\documentclass[]{pasj01}
\usepackage{natbib,color}

\begin{document} 
\Received{}
\Accepted{}

\title{Two- and three-dimensional wide-field weak lensing mass maps from the
  Hyper Suprime-Cam Subaru Strategic Program S16A data}

\author{Masamune \textsc{Oguri}\altaffilmark{1,2,3}}%
\author{Satoshi \textsc{Miyazaki}\altaffilmark{4,5}}%
\author{Chiaki \textsc{Hikage}\altaffilmark{3}}%
\author{\\Rachel \textsc{Mandelbaum}\altaffilmark{6}}%
\author{Yousuke \textsc{Utsumi}\altaffilmark{7}}%
\author{Hironao \textsc{Miyatake}\altaffilmark{8,3}}%
\author{\\Masahiro \textsc{Takada}\altaffilmark{3}}%
\author{Robert \textsc{Armstrong}\altaffilmark{9}}%
\author{James \textsc{Bosch}\altaffilmark{9}}%
\author{\\Yutaka \textsc{Komiyama}\altaffilmark{4,5}}
\author{Alexie \textsc{Leauthaud}\altaffilmark{10}}%
\author{Surhud \textsc{More}\altaffilmark{3}}
\author{\\Atsushi J. \textsc{Nishizawa}\altaffilmark{11}}
\author{Nobuhiro \textsc{Okabe}\altaffilmark{7,12,13}}%
\author{Masayuki \textsc{Tanaka}\altaffilmark{4}}

\altaffiltext{1}{Research Center for the Early Universe, University of Tokyo, Tokyo 113-0033, Japan}
\altaffiltext{2}{Department of Physics, University of Tokyo, Tokyo 113-0033, Japan}
\altaffiltext{3}{Kavli Institute for the Physics and Mathematics of the Universe (Kavli IPMU, WPI), University of Tokyo, Chiba 277-8582, Japan}
\altaffiltext{4}{National Astronomical Observatory of Japan, Mitaka, Tokyo 181-8588, Japan}
\altaffiltext{5}{Department of Astronomy, School of Science, Graduate University for Advanced Studies, Mitaka, Tokyo 181-8588, Japan}
\altaffiltext{6}{McWilliams Center for Cosmology, Department of Physics, Carnegie Mellon University, Pittsburgh, PA 15213, USA}
\altaffiltext{7}{Hiroshima Astrophysical Science Center, Hiroshima University, Higashi-Hiroshima, Kagamiyama 1-3-1, 739-8526, Japan}
\altaffiltext{8}{Jet Propulsion Laboratory, California Institute of Technology, Pasadena, CA 91109, USA}
\altaffiltext{9}{Department of Astrophysical Sciences, Princeton University, 4 Ivy Lane, Princeton, NJ 08544, USA}
\altaffiltext{10}{Department of Astronomy and Astrophysics, University of California, Santa Cruz, 1156 High Street, Santa Cruz, CA 95064, USA}
\altaffiltext{11}{Institute for Advanced Research, Nagoya University, Nagoya 464-8602, Aichi, Japan}
\altaffiltext{12}{Department of Physical Science, Hiroshima University, 1-3-1 Kagamiyama, Higashi-Hiroshima, Hiroshima 739-8526, Japan}
\altaffiltext{13}{Core Research for Energetic Universe, Hiroshima University, 1-3-1, Kagamiyama, Higashi-Hiroshima, Hiroshima 739-8526, Japan}

\email{masamune.oguri@ipmu.jp}





\maketitle

\begin{abstract}
We present wide-field (167~deg$^2$) weak lensing mass maps from the Hyper
Supreme-Cam Subaru Strategic Program (HSC-SSP).  We compare these weak
lensing based dark matter maps with maps of the distribution of the
stellar mass associated with  luminous red galaxies.  We find a strong
correlation between these two maps with a  correlation coefficient of
$\rho=0.54\pm0.03$ (for a smoothing size of $8'$).  This correlation
is detected even with a smaller smoothing scale of $2'$
($\rho=0.34\pm 0.01$). This detection is made uniquely possible
because of the high source density of the HSC-SSP weak lensing survey
($\bar{n}\sim 25$~arcmin$^{-2}$). We also present a variety of tests
to demonstrate that our maps are not significantly affected by
systematic effects. By using the photometric redshift information
associated with source galaxies, we reconstruct a three-dimensional
mass map. This three-dimensional mass map is also found to correlate
with the three-dimensional galaxy mass map. Cross-correlation tests
presented in this paper demonstrate that the HSC-SSP weak lensing mass
maps are ready for further science analyses.  
\end{abstract}

\section{Introduction}
Weak gravitational lensing has developed into a useful probe of the
Universe. Weak gravitational lensing takes advantage of small
distortions (``shears'') of distant galaxies which are caused by
the deflection of light ray paths by intervening gravitational
fields, as predicted by General Relativity. While the matter component
of the Universe is dominated by dark matter which cannot directly be
seen, weak gravitational lensing allows us to directly map out the
total mass distribution including dark matter.  

Many weak lensing studies focus on two-point statistics of the shear,
which is a direct observable of weak lensing analyses, such as cosmic
shear and tangential shear around galaxies and clusters of galaxies
\citep[e.g.,][]{bartelmann01,hoekstra08}. However, one can also
directly reconstruct the projected mass distribution from the observed
shear map \citep[e.g.,][]{kaiser93,seitz95,schneider96}. Such weak
lensing mass maps provide an important means of studying the
large-scale structure of the Universe as well as non-Gaussian features
of the matter density field. For example, massive clusters of galaxies
can be identified from peaks in mass maps
\citep[e.g.,][]{miyazaki02,miyazaki07,wittman06,shan12,utsumi14,liu15}.  
Correlations of mass maps with light maps constructed from galaxies
and clusters of galaxies can reveal the connection between mass and
light, by constraining mass-to-light ratios and galaxy biases 
\citep[e.g.,][]{hoekstra02,okabe10,jullo12,chang16,pujol16,utsumi16}. 
These applications of mass maps may be enhanced
further by interpolation methods to recover mass maps in masked
regions \citep[e.g.,][]{pires09,vanderplas12}.

Weak lensing mass maps have been constructed in many surveys,
including the Cosmological Evolution Survey
\citep[COSMOS;][]{massey07}, Deep Lens Survey \citep[DLS;][]{kubo09},
the Canada-France-Hawaii Telescope Lensing Survey
\citep[CFHTLenS;][]{vanwaerbeke13},
the CFHT/MegaCam Stripe-82 Survey \citep[CS82;][]{shan14},
Dark Energy Survey Science Verification data
\citep[DES SV;][]{chang15,vikram15},  the Kilo-Degree Survey
\citep[KiDS;][]{kuijken15}, and the Red Cluster Sequence 
Lensing Survey \citep[RCSLenS;][]{hildebrandt16}. It has been shown
that these mass maps correlate well with the light distributions
estimated from galaxies, which demonstrated the power of weak lensing
measurements for these surveys. These mass maps have also been used to
check residual systematics, by cross-correlating them with any
parameters related to Point Spread Function (PSF) and observing
conditions, as residual systematics in weak lensing measurements
can produce apparent correlations with these parameters.

While weak gravitational lensing for a fixed source redshift can only
probe the two-dimensional matter density field projected along the
line-of-sight, one can reconstruct the three-dimensional mass
distribution as well by combining weak lensing mass reconstructions in
different source redshift bins 
\citep[e.g.,][]{taylor01,hu02,bacon03,simon09,vanderplas11,leonard14,bohm17}. 
The idea has been applied to observations to obtain three-dimensional
mass maps \citep{taylor04,massey07,simon12}, although they have been
restricted to relatively small areas given the requirement of high
source galaxy densities. One can improve the accuracy of
three-dimensional mass reconstructions by using some information from
galaxy distributions as a prior \citep[e.g.,][]{amara12,szepietowski14},
although with this approach the resulting mass maps are no longer
independent of galaxy light distributions.

In this paper, we present two-dimensional and three-dimensional mass
maps from the Hyper Suprime-Cam (HSC) Subaru Strategic Program
\citep{aihara17a,aihara17b}, a wide-field imaging survey using the
Hyper Suprime-Cam \citep{miyazaki17a} mounted on the Subaru
8.2-meter telescope. Weak lensing analysis of commissioning data has
already demonstrated that the HSC is a powerful 
instrument for weak lensing studies \citep{miyazaki15,utsumi16}. 
The purpose of this paper is to construct weak lensing mass maps
to check the performance of weak lensing measurements in the HSC survey.
To do so, we cross-correlate weak lensing mass maps with the
distribution of stellar masses of red galaxies, which are known to
trace the large-scale structure of the Universe. We also
cross-correlate our mass maps with maps of various quantities such
as PSF and seeing sizes, to check for any possible
residual systematics in the reconstructed mass maps. Validating mass
maps is important for future applications of weak lensing mass maps in
the HSC survey, including the construction of mass-selected cluster
samples and cross-correlations of weak lensing maps with other surveys
such as ACTPol. 

This paper is organized as follows. In Section~\ref{sec:data}, we
describe our source galaxy catalog for weak lensing analysis as well
as our photometric red galaxy sample used for constructing galaxy mass
maps. Our two-dimensional mass map analysis is presented in
Section~\ref{sec:2dmap}, whereas our three-dimensional mass map
analysis is presented in Section~\ref{sec:3dmap}. We summarize our
result in Section~\ref{sec:summary}. Unless otherwise specified, we
assume a flat $\Lambda$-dominated cosmology with the matter density
$\Omega_M=0.27$, the baryon density $\Omega_b=0.045$, cosmological
constant $\Omega_\Lambda=0.73$, dimensionless Hubble constant
$h=0.71$, the power spectrum tilt $n_s=0.96$, and the normalization of
the density fluctuation $\sigma_8=0.80$.
We note that our conclusion is insensitive to the
choice of the cosmological parameters.

\section{Data}\label{sec:data}

\subsection{Weak lensing shear catalog}\label{sec:shearcat}

Galaxy shape measurements and the resulting shear catalog in the HSC
S16A dataset are detailed in \citet{mandelbaum17}. In short, the 
shapes of galaxies in the coadded $i$-band images are estimated using
the re-Gaussianization method \citep{hirata03}, and are calibrated using
simulated galaxy images that are similar to those used in GREAT3
\citep{mandelbaum15}. The image simulation includes
  realistic HSC PSFs and is carefully designed to reproduce the
  observed distribution of galaxy properties remarkably well, which
  allows a reliable estimate shear calibration and additive biases 
  (see Mandelbaum et al., in prep.).
We use conservative cuts for selecting galaxies
with secure shape measurements, e.g., $S/N\geq 10$ and $i\leq 24.5$.
The shear catalog has been tested and shown to pass requirements for
cosmological studies. The shape catalog contains $\sim 12$~million
galaxies selected from 137~deg$^2$, giving an average raw
number density of galaxies $\bar{n}\sim 25$~arcmin$^{-2}$.

The HSC S16A dataset consists (mostly) of 6 patches; XMM, GAMA09H,
GAMA15H, HECTOMAP, VVDS, and WIDE12H. While we present mass maps for
these individual patches separately, we combine our results on
cross-correlations for all these 6 patches. 

Accurate photometric redshifts for the shape catalog are important,
particularly for three-dimensional weak lensing mass reconstructions.
Thus we apply an additional cut to select galaxies with secure
photometric redshifts. We do so by selecting galaxies with the 
standard deviation computed from the probability distribution function
(PDF) of the photometric redshift smaller than 0.3. This cut removes
$\sim 16$\% of the galaxies from the shape catalog. While photometric
redshifts are measured for the HSC galaxies using several different
techniques, throughout this paper we use the {\tt mlz} photometric
redshifts \citep{tanaka17}. 

We use mock shear catalogs to estimate statistical uncertainties on
the mass maps. Details of the mock shear catalogs are given in
Appendix~1. 

\subsection{Galaxy catalog}\label{sec:galcat}

We need a galaxy sample with reasonably accurate redshift information
in order to compare mass maps from weak lensing with galaxy
distributions. While the HSC survey footprint overlaps SDSS, the
redshift coverage of SDSS spectroscopic galaxies is limited. Following
redMaGiC \citep{rozo16}, in this paper we construct a photometrically 
selected sample of luminous red galaxies (LRGs) from the HSC data by
taking advantage of the Stellar Population Synthesis (SPS) fitting method
developed for the CAMIRA algorithm \citep{oguri14,oguri17}.
The CAMIRA algorithm fits all galaxies with the SPS model of passively
evolving galaxies from \citet{bruzual03}, with careful corrections for
slight color differences between the model and observations using
spectroscopic  galaxies, to compute the goodness-of-fit $\chi^2$ which
is used to construct a three-dimensional richness map for identifying
clusters of galaxies. The calibration for the HSC survey and the
resulting cluster catalog is presented in \citet{oguri17}, in which
$\sim 2000$ clusters of galaxies at $0.1<z<1.1$ selected from the area
of $\sim 230$~deg$^2$ are reported.

We use this SPS model calibrated in the HSC survey
\citep[see][]{oguri17} to select LRGs as follows. We fit all galaxies
with the SPS model, leaving redshift, stellar mass, and
metallicity as model parameters. In this model a single instantaneous
burst at the formation redshift $z_f=3$ is assumed, and a prior is
added to the metallicity \citep[see][]{oguri14}. Since we only consider
passively evolving galaxies in the SPS model, any galaxies that can be
fit well with the SPS model are red galaxies. Specifically, we select
galaxies with best-fit $\chi^2<10$ (3 degrees of freedom). In order to
construct a roughly volume-limited galaxy sample, we restrict the
redshift range to $0.05<z_{\rm photo}<1.05$, where $z_{\rm photo}$ is
the best-fit photometric redshift, and the stellar mass range to
$M_*>10^{10.3}M_\odot$, where the stellar mass is derived assuming the
\citet{salpeter55} initial mass function. From the HSC S16A Wide
dataset, we select 1,024,729 LRGs that satisfy these criteria. 

\begin{figure}
 \begin{center}
  \includegraphics[width=8cm]{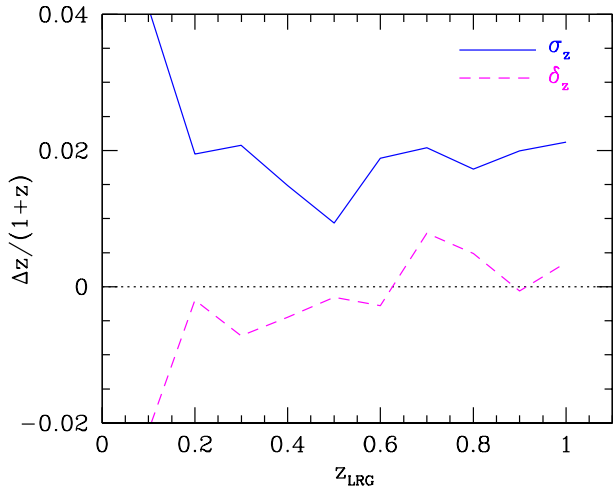} 
 \end{center}
\caption{Comparison of photometric redshifts of LRGs $z_{\rm LRG}$ with
  their spectroscopic redshifts $z_{\rm spec}$. We plot the scatter
  $\sigma_z$ ({\it solid}) and bias $\delta_z$ ({\it dashed}) of the
  residual $(z_{\rm LRG}-z_{\rm spec})/(1+z_{\rm spec})$ as a
  function of redshift.}
\label{fig:lrg_zcomp} 
\end{figure}

\begin{figure}
 \begin{center}
  \includegraphics[width=8cm]{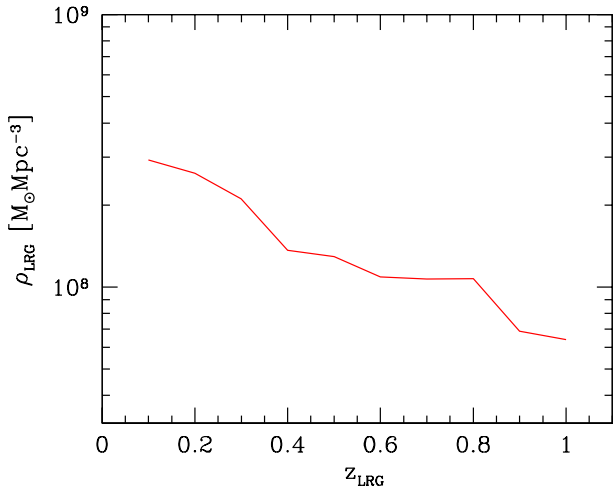} 
 \end{center}
\caption{Stellar mass densities of LRGs $\rho_{\rm LRG}$ as a function
  of photometric redshift, for LRGs with stellar masses
  $M_*>10^{10.3}M_\odot$. }\label{fig:lrg_ngal}  
\end{figure}

We cross match the LRGs with spectroscopic galaxies in the HSC
footprint \citep[see][and references therein]{oguri17} to check the
accuracy of the photometric redshifts of LRGs. Using 51,402 LRGs that
match the spectroscopic galaxy catalog, we derive the scatter
$\sigma_z$ and bias $\delta_z$ of the residual $(z_{\rm LRG}-z_{\rm
  spec})/(1+z_{\rm spec})$. Figure~\ref{fig:lrg_zcomp} shows the
scatter and bias as a function of redshift. Here we apply $3\sigma$
clipping when estimating the scatter and bias. The resulting outlier
rate is $\sim 7\%$ for the whole sample. We find that the scatter
is $\sigma_z\sim 0.02$ for the whole redshift range of interest,
except for the lowest redshift $z\sim 0.1$ which is probably due to
relatively poor photometric accuracy of nearby, very bright galaxies in
the HSC survey. The relatively poor photometric redshifts at the
lowest redshift may also be due to the lack of $u$-band images.
We also note that the scatter is larger than the
scatter of photometric redshifts of CAMIRA clusters, $\sigma_z\lesssim
0.01$, because the cluster photometric redshifts are derived by
combining photometric redshifts of several cluster member galaxies.

In Figure~\ref{fig:lrg_ngal}, we derive the stellar mass density of
LRGs by summing up the stellar masses of all the LRGs with $M_*>10^{10.3}M_\odot$.
The stellar mass density increases from $z=1$ to $0$, which is broadly
consistent with previous analysis of the evolution of early-type
galaxies \citep[e.g.,][]{bell04}.

\begin{figure*}
 \begin{center}
  \includegraphics[width=11.2cm]{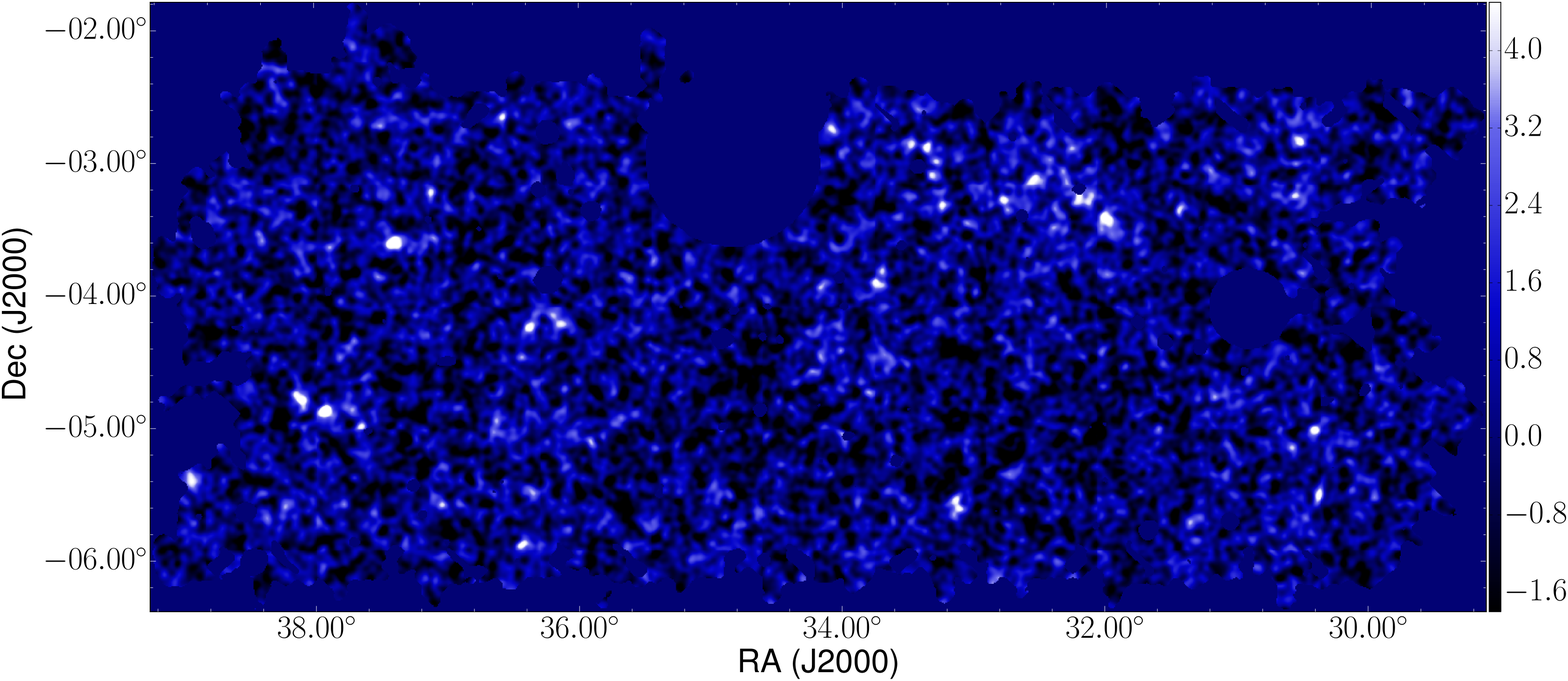} \\
\hspace*{2.5mm}  \includegraphics[width=11.5cm]{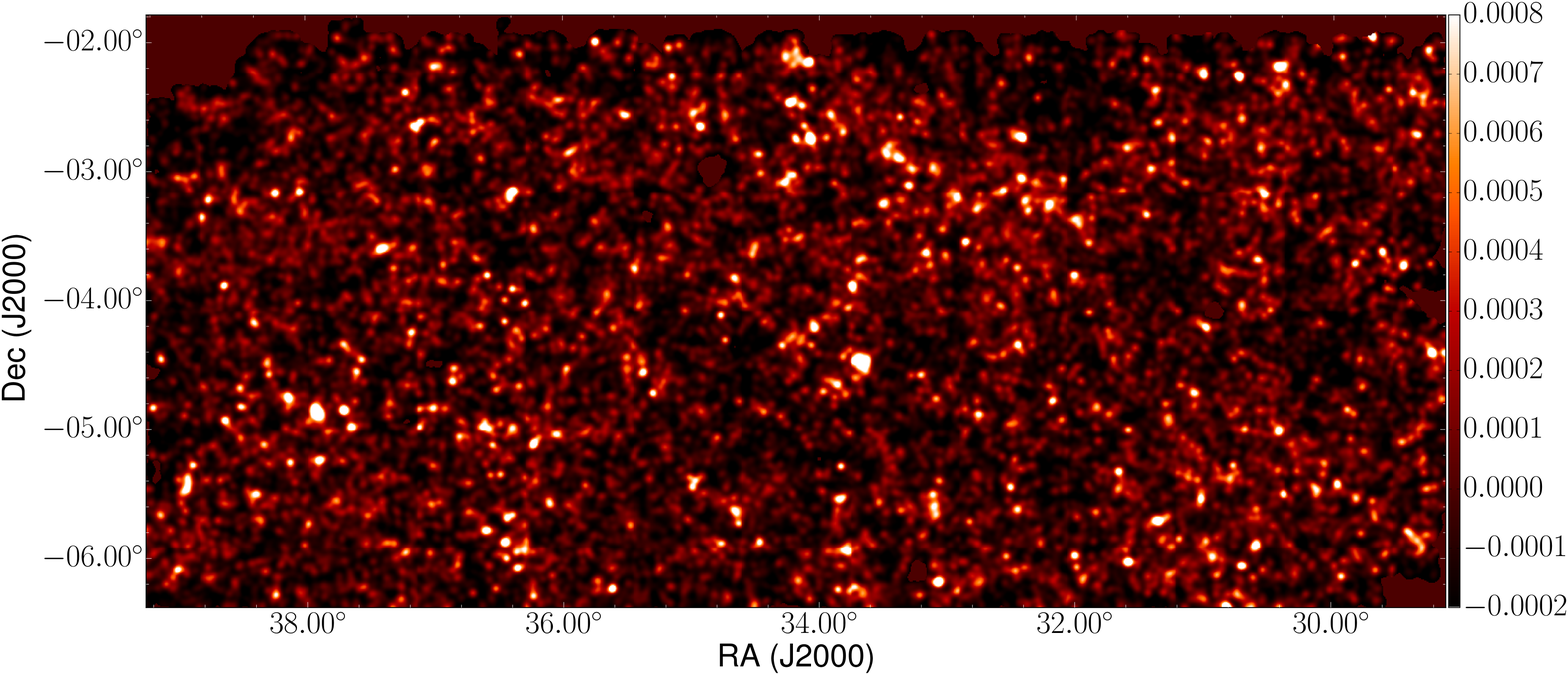} 
 \end{center}
\caption{Total mass ({\it upper}) and galaxy mass ({\it lower}) maps in the
  XMM field. In the total mass map, we show the S/N of the
  weak lensing reconstructed map, which is roughly proportional to the
  convergence $\kappa$. In the galaxy mass map we directly show the
  galaxy mass map value $\kappa_{\rm g}$ defined in
  equation~(\ref{eq:kappag}). The smoothing scale is $\theta_s=2'$
  (see equation~\ref{eq:Gaussian}). }
\label{fig:map_XMM} 
\end{figure*}

\begin{figure*}
 \begin{center}
  \includegraphics[width=13.5cm]{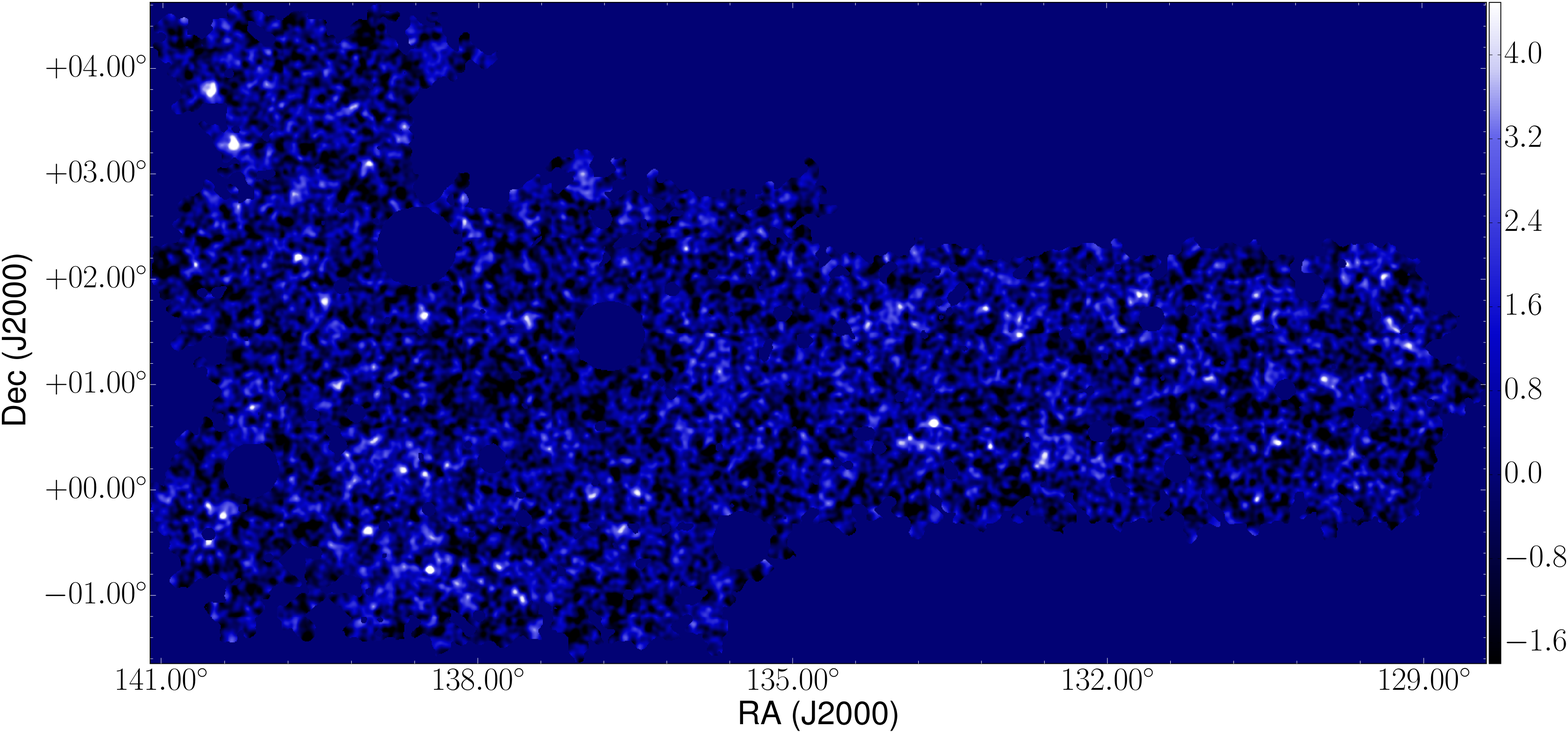} \\
\hspace*{2.5mm} \includegraphics[width=13.9cm]{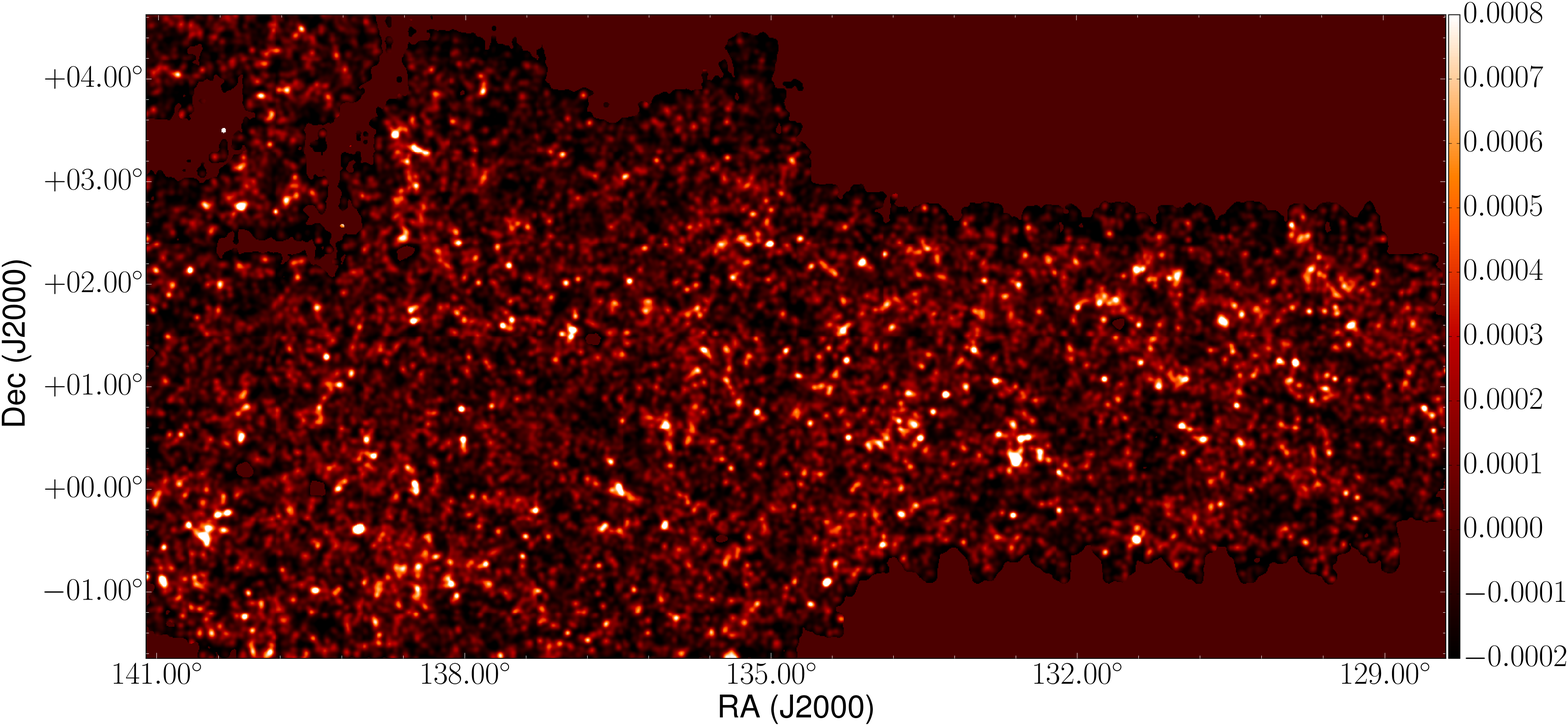} 
 \end{center}
\caption{Same as Figure~\ref{fig:map_XMM}, but for the GAMA09H field.}
\label{fig:map_G09} 
\end{figure*}

\begin{figure*}
 \begin{center}
  \includegraphics[width=8.3cm]{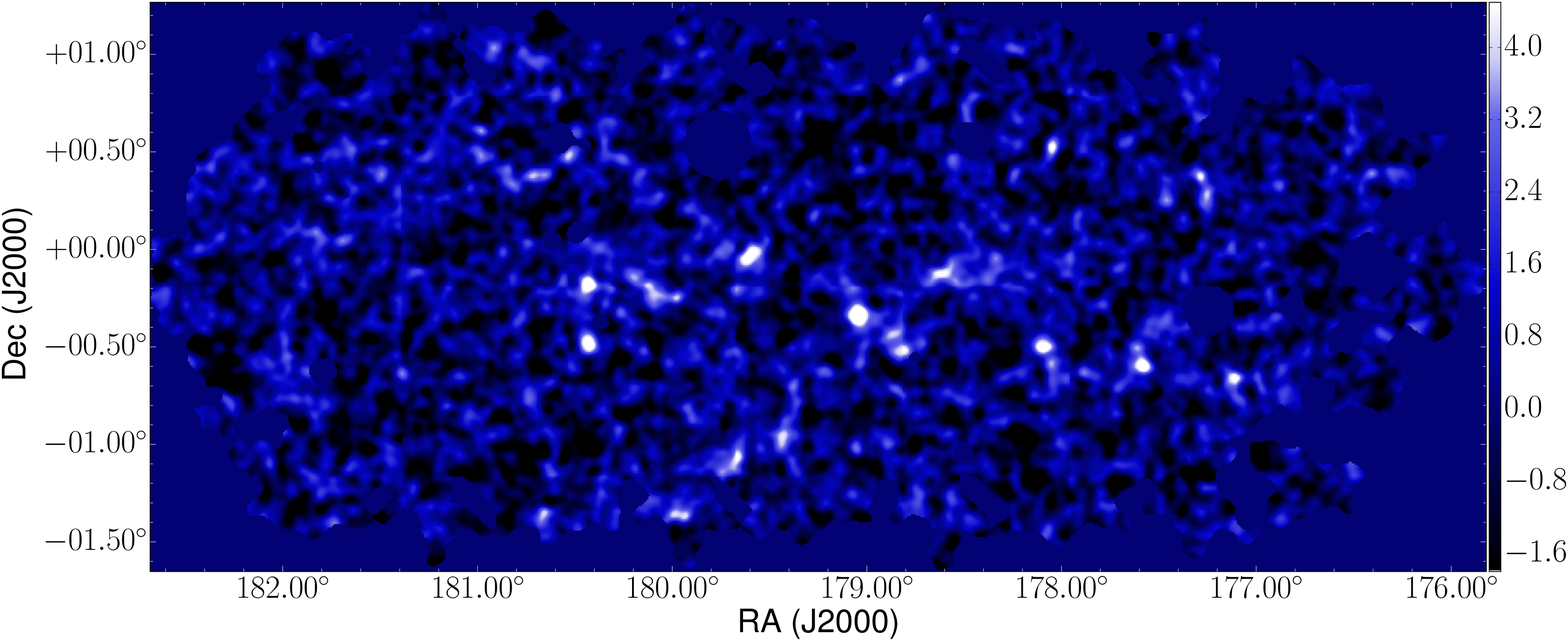} \\
\hspace*{2.1mm} \includegraphics[width=8.55cm]{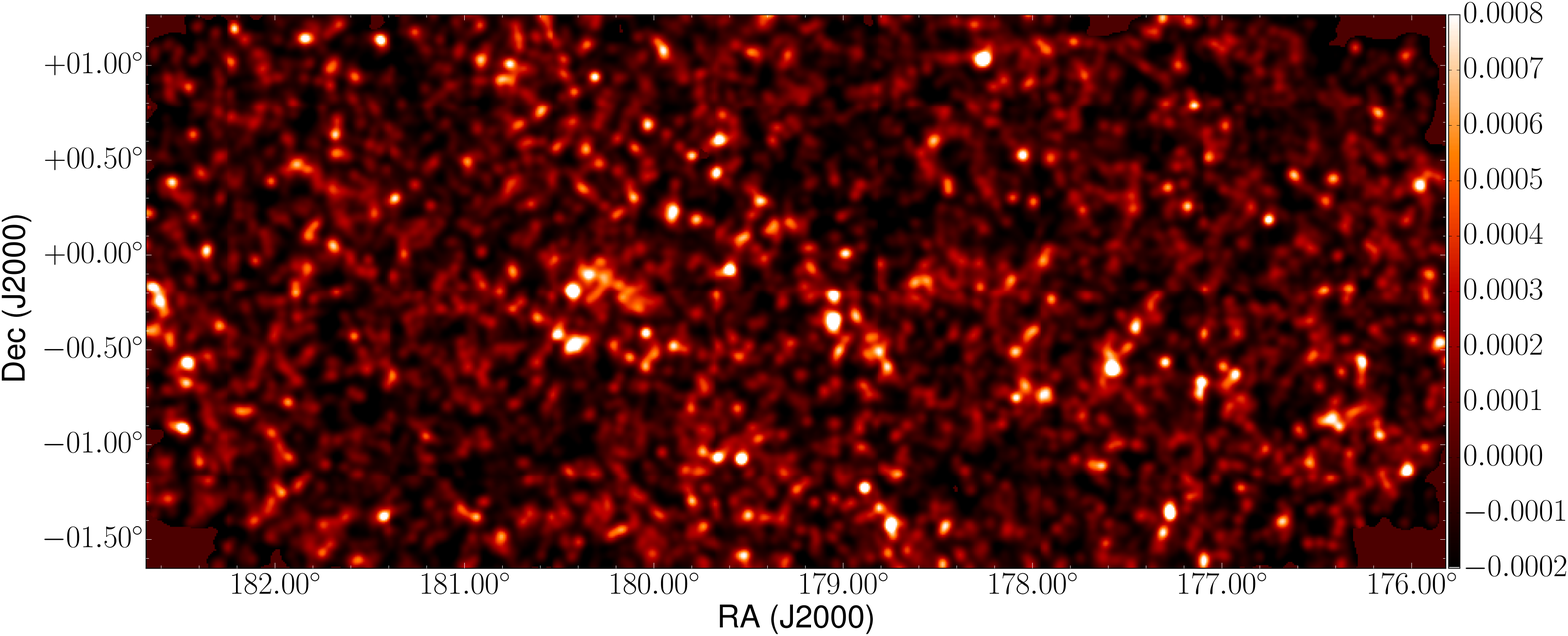} 
 \end{center}
\caption{Same as Figure~\ref{fig:map_XMM}, but for the WIDE12H field.}
\label{fig:map_W12} 
\end{figure*}

\begin{figure*}
 \begin{center}
  \includegraphics[width=13.5cm]{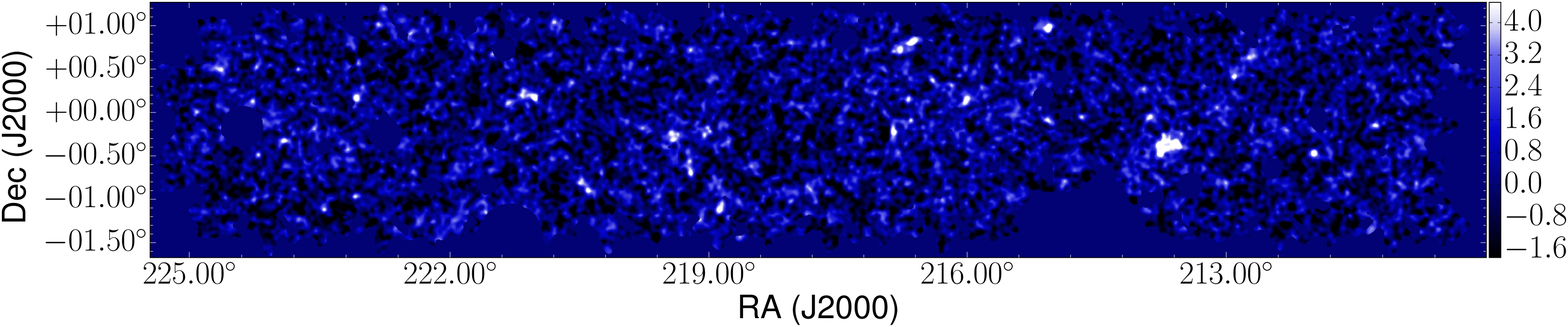}  \\
 \hspace*{2.5mm}  \includegraphics[width=13.9cm]{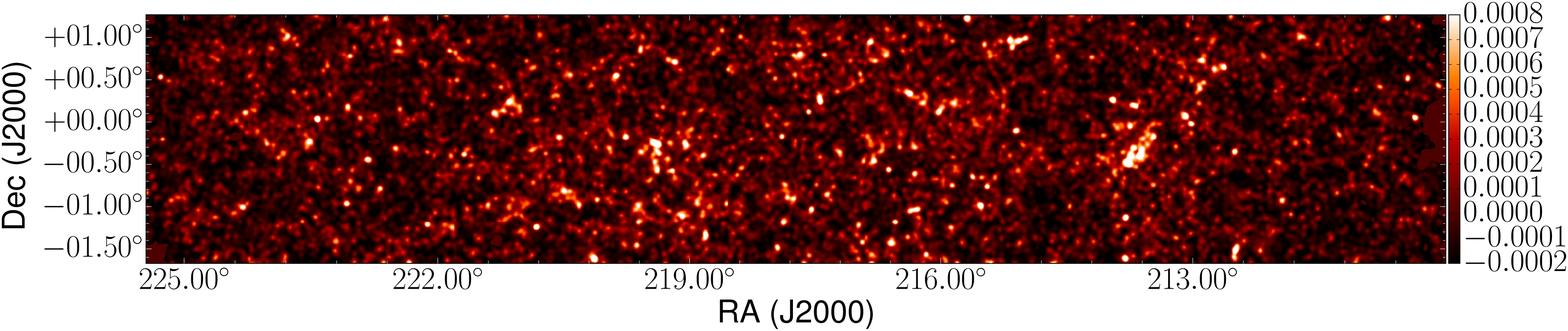} 
 \end{center}
\caption{Same as Figure~\ref{fig:map_XMM}, but for the GAMA15H field.}
\label{fig:map_G15} 
\end{figure*}

\begin{figure*}
 \begin{center}
  \includegraphics[width=9.0cm]{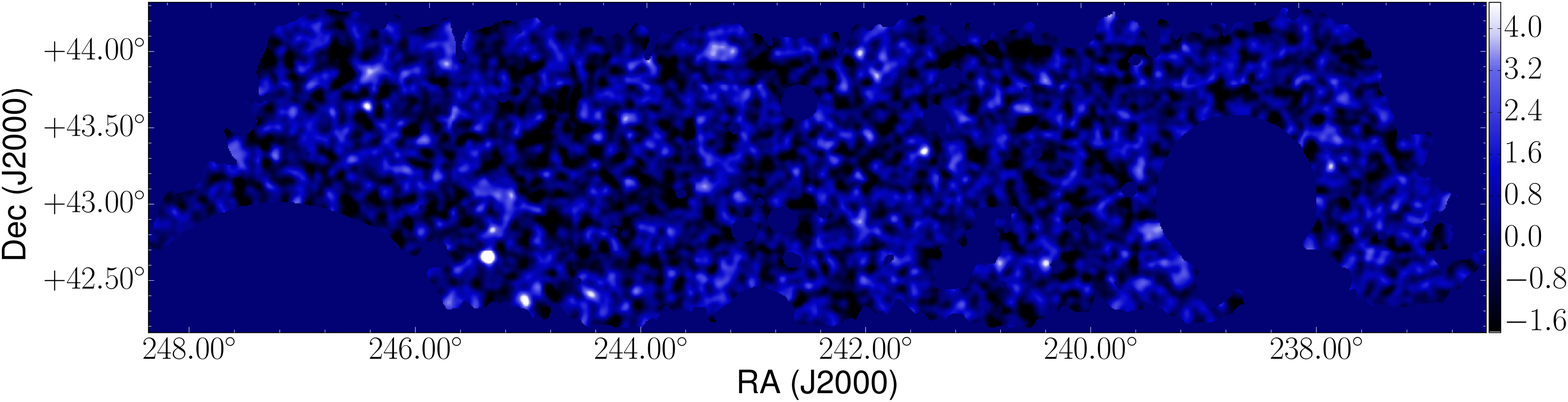}  \\
 \hspace*{2.1mm} \includegraphics[width=9.25cm]{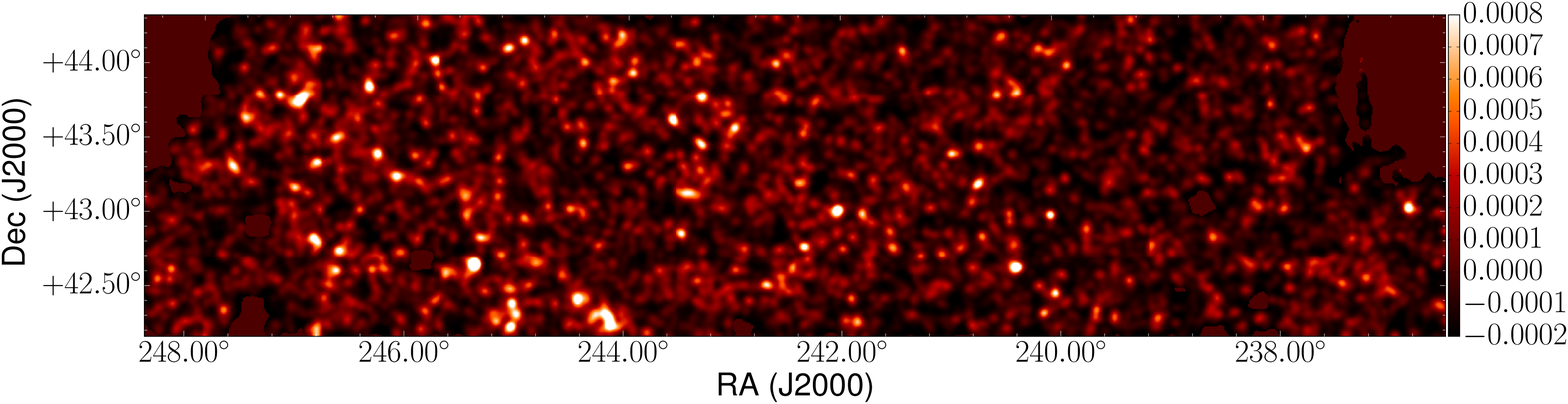} 
 \end{center}
\caption{Same as Figure~\ref{fig:map_XMM}, but for the HECTOMAP field.}
\label{fig:map_HEC} 
\end{figure*}

\begin{figure*}
 \begin{center}
  \includegraphics[width=12cm]{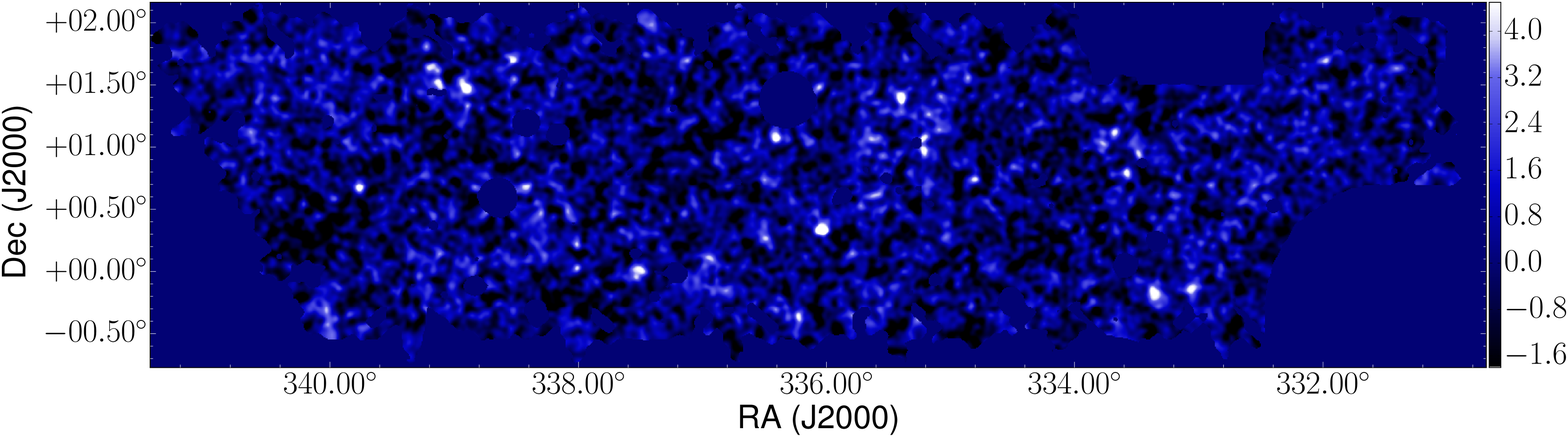} \\
 \hspace*{2.5mm} \includegraphics[width=12.3cm]{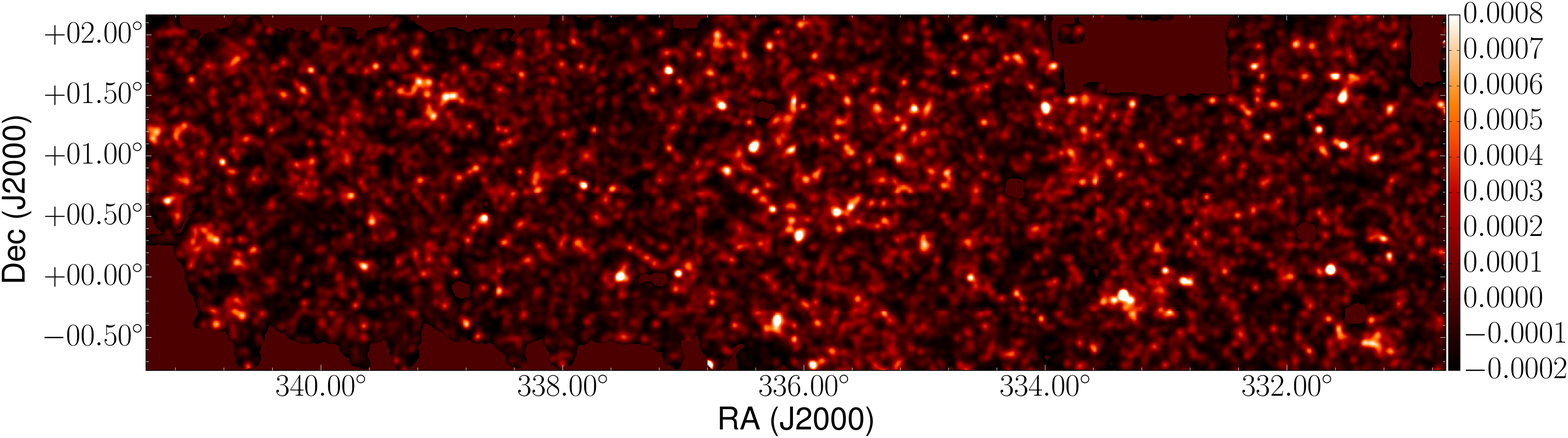} 
 \end{center}
\caption{Same as Figure~\ref{fig:map_XMM}, but for the VVDS field.}
\label{fig:map_VVD} 
\end{figure*}

\section{Two-dimensional mass maps}\label{sec:2dmap}

\subsection{Mass reconstruction technique}\label{sec:wlmap2d}

The construction of mass maps from weak lensing requires spatial
filtering to reduce noise. There are several possible choices of
spatial filters which must be chosen depending on the application
of the mass maps. \citet{miyazaki17b} also constructs
wide-field mass maps from the same HSC data, but they are interested
in identifying clusters of galaxies from peaks in mass maps. In
searching for clusters, it is beneficial to use spatial filters
that eliminate the large-scale power in order to reduce the scatter in 
peak heights coming from the large-scale structure. In contrast,
since we are interested in the large-scale structure, in this paper we
use a Gaussian filter which retains the large-scale power. 
Systematic tests with $B$-mode mass maps are also presented in
\citet{mandelbaum17} and \citet{miyazaki17b}. 

In this paper we follow a mass reconstruction method proposed by
\citet{kaiser93}. Since we are interested in large-scale mass
distributions, in this paper we always consider the weak lensing
limit, $|\kappa| \ll 1$. First we smooth the shear field 
$\gamma_\alpha(\boldsymbol{\theta})$ ($\alpha=1$, $2$) as 
\citep{seitz95}
\begin{equation}
\hat{\gamma}_\alpha(\boldsymbol{\theta})=
\frac{\sum_i w_i\left[\gamma_\alpha
  (\boldsymbol{\theta}_i)-c_{\alpha,i}\right] W(|\boldsymbol{\theta}-\boldsymbol{\theta}_i|)}
{\sum_i w_i (1+m_i)W(|\boldsymbol{\theta}-\boldsymbol{\theta}_i|)},
\end{equation}
where $w_i$ is the inverse variance weight for the $i$-th
  galaxy given in the weak lensing shear catalog, 
the shear $\gamma_\alpha(\boldsymbol{\theta}_i)$ is related to
the distortion $e_\alpha$ 
(the ellipticity defined by second moments of the galaxy image)
as $\gamma_\alpha(\boldsymbol{\theta}_i)=e_\alpha(\boldsymbol{\theta}_i)/2{\cal
  R}$ with ${\cal R}$ being the shear responsivity
that connects the distortion and the shear, $W(\theta)$ is the
Gaussian smoothing kernel
\begin{equation}
W(\theta)=\frac{1}{\pi \theta_{\rm s}^2}\exp
\left(-\frac{\theta^2}{\theta_{\rm s}^2}\right),
\label{eq:Gaussian}
\end{equation}
and $m_i$ and $c_{\alpha,i}$ are the multiplicative and additive biases
for the $i$-th galaxy (see \citealt{mandelbaum17} for more details on
the shear responsivity and calibration factors). 
We then convert the shear field to the convergence field via
\begin{equation}
\hat{\kappa}(\boldsymbol{\theta})=\frac{1}{\pi}\int d^2\theta' 
\frac{\hat{\gamma}_t(\boldsymbol{\theta'}|\boldsymbol{\theta})}
{|\boldsymbol{\theta}-\boldsymbol{\theta'}|^2},
\end{equation}
where $ \gamma_t(\boldsymbol{\theta'}|\boldsymbol{\theta})$ is a
tangential shear at position $\boldsymbol{\theta'}$ computed with
respect to the reference position $\boldsymbol{\theta}$. 

In practice we construct the mass map on a regular grid adopting a
flat-sky approximation. First, we create a pixelized shear map for each
of the 6 patches with a pixel size of $0\farcm5$, apply the Fast
Fourier Transform (FFT), and conduct the convolutions in the Fourier
space to obtain the smoothed convergence map, which is sometimes
referred as an $E$-mode mass map. Since FFT assumes a periodic
boundary condition, we apply zero padding beyond the boundary of the
image before FFT. The imaginary part of the reconstructed convergence
map represents a $B$-mode mass map, which is used to check for certain
types of residual systematics in our weak lensing measurements. In
\citet{mandelbaum17}, we show that the $B$-mode mass map PDF well
follows the Gaussian distribution as expected for weak lensing mass
maps without significant systematic errors in shape measurements. In
fact, the boundary effect induces small non-vanishing $B$-mode
signals, which can be estimated from our mock shape catalog which has
exactly the same geometry as our input HSC shape catalog \citep[see
  also][]{mandelbaum17}.  

We also construct a noise map as follows. We randomly rotate the
orientations of individual galaxies, and construct a mass map using
the randomized galaxy catalog. We repeat this procedure to create
300 random mass maps from 300 realizations of randomized galaxy
catalogs. We then compute a standard deviation of each pixel from the
300 random mass maps to construct a ``sigma map'', a map showing the
spatial variation of the statistical noise of the reconstructed mass
map. The sigma map includes only the shape noise and measurement
error, and does not include cosmic shear. From the sigma map we can
define the signal-to-noise ratio (S/N) for each pixel simply from the
ratio of the $\kappa$ value of the reconstructed mass map to the
standard deviation of $\kappa$ from the sigma map. 

In real observations, there are several regions where data are
missing due to bright star masks and edges. Reconstructed mass maps in
and near those regions are noisy and are not suitable for analysis.
To determine the mask region for each mass map, we construct a number
density map of the input galaxy catalog by convolving the number
density in each pixel with the same smoothing kernel which was used in
constructing mass maps (equation~\ref{eq:Gaussian}). Then we derive the
mean of the number density map with 2.5$\sigma$ clipping. We adopt
clipping because the number density map has a non-Gaussian tail. 
We mask all pixels with the {\it smoothed} number density less than
0.5 times the mean number density computed above, assuming that they
correspond to edges and regions that are affected by bright star
masks. In addition, we derive the mean of the sigma map with
2.5$\sigma$ clipping and mask all pixels with the noise value larger
than twice the mean value, although this additional cut removes only a
minor fraction of the survey area.
The criteria for these masking procedures are
  determined empirically so that the degradation of the mass map at
  the edge is not significant.

We show the mass maps of the 6 HSC S16A patches in
Figures~\ref{fig:map_XMM}, \ref{fig:map_G09}, \ref{fig:map_W12},
\ref{fig:map_G15}, \ref{fig:map_HEC}, and \ref{fig:map_VVD}.
These mass maps are created using a relatively small smoothing scale
of $\theta_s=2'$. Here we show S/N maps which are similar to $\kappa$
maps except near the edges where the noise is slightly larger. In the
cross-correlation analysis below we use $\kappa$ maps rather than S/N maps.
The total area of unmasked regions in these mass maps is $\sim
167$~deg$^2$, which is larger than the total area of the regions where
the weak lensing shape catalog is defined, $\sim 137$~deg$^2$
\citep[see][]{mandelbaum17}, because of the non-local nature of the
weak lensing mass reconstruction.

\subsection{Galaxy mass maps}\label{sec:galmap2d}

The LRG sample constructed in Section~\ref{sec:galcat} is used to
create a galaxy mass map, a projected map of stellar masses of LRGs
with the same redshift weight as weak lensing. Specifically, we
compute a galaxy mass map value in each pixel as
\begin{equation}
\hat{\kappa}_{\rm
  g}(\boldsymbol{\theta}_i)=
\sum_k\frac{M_{*,k}}{(D(z_k)\Delta\theta)^2\Sigma_{\rm crit}(z_k)},
\label{eq:kappag}
\end{equation}
where $k$ runs over LRGs that fall within a pixel centered at
$\boldsymbol{\theta}_i$, $M_{*,k}$ is the stellar mass of $k$-th LRG,
$D(z_k)$ is the angular diameter distance to the LRG photometric
redshift $z_k$, and $\Delta\theta=0\farcm5$ is the size of each
pixel. The critical surface density $\Sigma_{\rm crit}^{-1}(z_k)$ is
computed as 
\begin{equation}
  \Sigma_{\rm crit}^{-1}(z_k)=\frac{4\pi G}{c^2} D(z_k)\int_{z_k}^\infty dz\,p(z)
  \frac{D(z_k,z)}{D(z)},
\end{equation}
where $p(z)$ is the average PDF of photometric redshifts of source
galaxies used for the weak lensing analysis, and $D(z_k,z)$ and $D(z)$
are angular diameter distances from redshift $z_k$ to $z$ and from
redshift $0$ to $z$, respectively. Strictly speaking, the critical
surface density has some spatial variation from the large-scale
structure of source galaxies, which is not taken in account in the
following analysis. From equation~(\ref{eq:kappag}), we subtract the
mean value as $\hat{\kappa}_{\rm g}(\boldsymbol{\theta}_i)\rightarrow
\hat{\kappa}_{\rm g}(\boldsymbol{\theta}_i)-\bar{\hat{\kappa}}_{\rm
  g}$, and apply the same Gaussian smoothing kernel
(equation~\ref{eq:Gaussian}) as used for the weak lensing mass map to
obtain the final galaxy mass map. 

We show the galaxy mass maps of the 6 HSC S16A patches in
Figures~\ref{fig:map_XMM}, \ref{fig:map_G09}, \ref{fig:map_W12},
\ref{fig:map_G15}, \ref{fig:map_HEC}, and \ref{fig:map_VVD}, which are
created using the same smoothing scale of $\theta_s=2'$ as for the
weak lensing mass maps.

\subsection{Cross-correlations of maps}\label{sec:cc2d}

We quantify the correlation between mass maps from weak lensing and
galaxy mass maps from photometric LRGs using the Pearson correlation
coefficient. For any two maps $\kappa_1(\boldsymbol{\theta}_i)$ and
$\kappa_2(\boldsymbol{\theta}_i)$ with zero means, $\langle
\kappa_i\rangle=0$, the correlation coefficient $\rho_{\kappa_1\kappa_2}$
is defined as
\begin{equation}
  \rho_{\kappa_1\kappa_2}
  =\frac{\sum_i\kappa_1(\boldsymbol{\theta}_i)\kappa_2(\boldsymbol{\theta}_i)}
  {\left[\sum_i\left\{\kappa_1(\boldsymbol{\theta}_i)\right\}^2\right]^{1/2}
    \left[\sum_i\left\{\kappa_2(\boldsymbol{\theta}_i)\right\}^2\right]^{1/2}},
\label{eq:pearson}
\end{equation}
where the summation runs over the pixels. The correlation coefficient
becomes $\rho_{\kappa_1\kappa_2}\sim 0$ if the two maps are
independent, whereas $\rho_{\kappa_1\kappa_2}\sim 1$ if the two maps
are highly correlated. 

\begin{figure}
 \begin{center}
  \includegraphics[width=8cm]{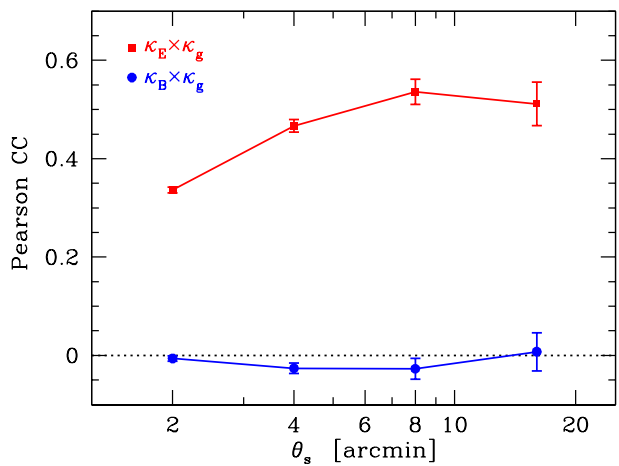} 
 \end{center}
\caption{Pearson correlation coefficients (equation~\ref{eq:pearson})
  between mass maps from weak lensing and galaxy mass maps from LRGs
  as a function of the smoothing size $\theta_s$ in
  equation~(\ref{eq:Gaussian}). Filled squares show cross-correlations
  between the $E$-mode mass map ($\kappa_{\rm E}$) and the galaxy mass
  map ($\kappa_{\rm g}$). Filled circles show the cross-correlation
  between the $B$-mode mass map ($\kappa_{\rm B}$) and the galaxy mass
  map ($\kappa_{\rm g}$). Errors are estimated from 50 mock samples
  of the weak lensing shear catalog, which include cosmic variance
  (see Appendix~1).}
\label{fig:plot_cormap_smoothing} 
\end{figure}

We cross-correlate $E$-mode and $B$-mode mass maps
(Section~\ref{sec:wlmap2d}) with galaxy mass maps
(Section~\ref{sec:galmap2d}). Since $E$-mode mass maps correspond to
the true matter distributions, we expect that the galaxy mass maps
correlate only with $E$-mode mass maps. 
Figure~\ref{fig:plot_cormap_smoothing} shows the correlation
coefficients as a function of the smoothing size $\theta_s$ in
the Gaussian smoothing kernel. Here we combine the cross-correlation
results of all the 6 HSC S16A patches. For each patch, we
compute the cross-correlation coefficients and estimate their errors
using the 50 mock samples of the weak lensing shape catalog
(Section~\ref{sec:shearcat}).
We use the standard deviation of cross-correlation
  coefficients for the 50 mock samples as our error estimate.
We then compute the inverse-variance
weighted average of the correlation coefficient of each map
combination which we show in Figure~\ref{fig:plot_cormap_smoothing}.

Figure~\ref{fig:plot_cormap_smoothing} indicates that the $E$-mode mass
maps indeed correlates well with the galaxy mass maps. The correlation
coefficients are consistent with zero for the $B$-mode mass maps.
We find that the correlation coefficients increase with increasing
smoothing size $\theta_s$, which is expected because larger smoothing
sizes reduce the statistical errors from the shot noise more
efficiently. It is worth noting that the HSC mass maps show
significant cross correlation ($\rho=0.34\pm 0.01$) even for the small
smoothing size of $\theta_s=2'$. This result should be compared with
previous wide-field mass maps constructed in CFHTLenS
\citep{vanwaerbeke13} and DES SV \citep{chang15,vikram15} for which
much larger smoothing sizes of $\theta_s\sim 7'$ are required to
obtain $\rho\sim 0.3$\footnote{Note that the definition of the
  smoothing sizes in this paper are different from these previous works
  by a factor of $\sqrt{2}$.}. This difference is
mainly due to the high density of the shape catalog for weak lensing
measurements in the HSC survey.
Our study demonstrates that the HSC
survey can generate mass maps at higher resolution than CFHTLenS and
DES SV, which is crucial for the construction of a mass-selected
cluster sample from weak lensing mass maps \citep{miyazaki17b}.   

Larger correlation coefficients with increasing smoothing scale is
understood as follows. The shot noise depends on the source number
density $\bar{n}$ and smoothing scale as $\sigma\propto
(\bar{n}\theta_s)^{-1}$, where in the range of our interest the
fluctuation of a smoothed mass map due to the large-scale structure
roughly scales as $\sigma_{\rm LSS}\propto \theta_s^{-0.4}$. Therefore
at large $\theta_s$ the shot noise becomes smaller than $\sigma_{\rm
  LSS}$ that produces a correlation between mass map and
 galaxy mass maps. This also suggests that the transition smoothing
 scale beyond which we see large correlation coefficients is inversely
 proportional to the source number density, which explains the
 difference between our results and previous results from CFHTLenS and
 DES SV. However, our result as well as previous results show that
 correlation coefficients do not approach to unity but saturate
 at $\sim 0.5-0.6$ at very large $\theta_s$, which is presumably due
 to the combination of several effects, including the limited redshift
 and mass ranges of the LRG sample, errors in the stellar mass and
 photometric redshift estimates, and the lack of blue galaxies in the
 galaxy sample. Intrinsic alignments may also affect our weak lensing
 mass maps, although the effect of intrinsic alignments on the
 correlation coefficients is expected to be relatively minor.

\subsection{Systematics tests}

\begin{figure}
 \begin{center}
  \includegraphics[width=8cm]{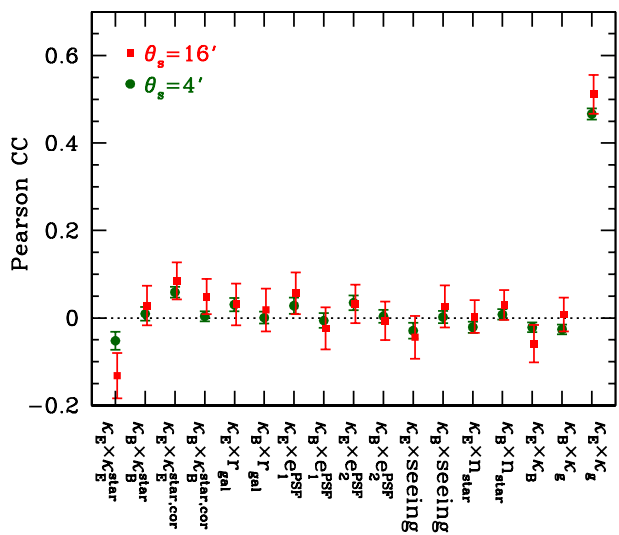} 
 \end{center}
\caption{Test of systematic effects in weak lensing mass maps from
  cross-correlations of mass maps with various quantities that are
  potentially a source of systematics \citep[see
    also][]{vikram15}. We show results for smoothing sizes of both
  $\theta_s=4'$ ({\it filled circles}) and $16'$ ({\it filled
    squares}). For comparison, the rightmost points show the
  cross-correlation coefficients between weak lensing and galaxy mass
  maps presented in Figure~\ref{fig:plot_cormap_smoothing}, which represents
  the physical cross correlation rather than the systematic test. 
  Errors are estimated from 50 mock samples of the weak lensing shear
  catalog, which include cosmic variance (see Appendix~1).} 
\label{fig:plot_cormap_all} 
\end{figure}

Following \citet{vikram15}, we also examine cross-correlations between
weak lensing mass maps and various maps using parameters that
potentially act as a source of systematic effects in generating mass
maps. While a number of tests have been performed in
\citet{mandelbaum17}, systematics tests based on the mass maps presented
here serve as additional checks for validating the shear catalog. 

The main source of systematics in weak lensing measurements comes from
the PSF. Imperfect corrections of PSFs in galaxy shape
measurements can generate artificial correlations between $E$- and
$B$-mode mass maps and PSF parameters. We construct star mass maps
$\kappa^{\rm star}_{\rm E}$ and $\kappa^{\rm star}_{\rm B}$ which use
star ellipticities $e_1^{\rm star}$ and $e_2^{\rm star}$ to construct
weak lensing mass maps with the same method as described in
Section~\ref{sec:wlmap2d}. The star catalog used for this analysis is
the same as the one used for various systematics tests in
\citet{mandelbaum17}. For this purpose, we use both the original 
star ellipticities $e_i^{\rm star}$ as well as star ellipticities after
the PSF correction is applied, i.e., $e_i^{\rm star,cor}=e_i^{\rm
  star}-e_i^{\rm PSF}$. We also create maps of star ellipticities
$e_1$ and $e_2$ themselves. These maps are 
constructed first by deriving their average values in each pixel and
convolve the maps of these average values with the Gaussian smoothing
kernel of equation~(\ref{eq:Gaussian}).

In addition, we create maps of seeing sizes, star densities $n_{\rm
  star}$, and average galaxy sizes of the shape catalog $r_{\rm gal}$,
as these parameters may also produce systematic effects in weak
lensing shape measurements. Again, these maps are smoothed with the
same smoothing kernel.

Figure~\ref{fig:plot_cormap_all} shows the results for smoothing sizes
of both $\theta_s=4'$ and $16'$. Again, results for all the 6 HSC S16A 
patches are combined. We find that cross-correlations between weak
lensing mass maps and the parameters considered above are small. All
the cross-correlations are consistent with zero within $\sim
2\sigma$ level (given the large number of cross-correlations
considered here, we naturally expect that some of the points can
deviate more than $1\sigma$ by chance), which is in marked contrast to
the cross-correlations between mass maps and galaxy mass maps, which
are detected quite significantly. A possible exception is
cross-correlations between star weak lensing mass maps and maps with
star (PSF) ellipticities, although their cross-correlation
coefficients are much smaller than the cross-correlations between mass
maps and galaxy mass maps. This small deviation from zero is
presumably due to small residual PSF leakage and PSF modeling errors
that are also seen in other systematics tests
\citep[see][]{mandelbaum17}. We conclude that our weak lensing mass
maps constructed in the HSC survey are not significantly affected by
systematic effects.   

\section{Three-dimensional mass maps}\label{sec:3dmap}

\subsection{Three-dimensional mass reconstruction}\label{sec:wlmap3d}

We can also reconstruct three-dimensional mass maps from weak lensing
by taking advantage of photometric redshift measurements for source
galaxies. We follow \citet{simon09} to use a linear algorithm with
the Wiener filtering for the three-dimensional mass reconstruction.

First we consider convergence $\kappa_l$ for the source redshift
bin $l$ at $z_{l,{\rm min}}<z<z_{l,{\rm max}}$. Since the convergence
is the projected matter density field, it can be described by a
weighted sum of the density fluctuation $\delta_k$ at redshift
$z_{k,{\rm min}}<z<z_{k,{\rm max}}$ as
\begin{eqnarray}
  \kappa_l &\approx &\sum_k \left[\int_{z_{k,{\rm min}}}^{z_{k,{\rm max}}} dz
  \frac{\bar{\rho}(z)}{H(z)(1+z)\Sigma_{{\rm
      crit},l}(z)}\right]\delta_k \nonumber\\
  &\equiv & \sum_k Q_{lk}\delta_k,
  \label{eq:del2kap}
\end{eqnarray}
where
$H(z)$ is the Hubble parameter at redshift $z$ and
the critical density $\Sigma_{{\rm crit},l}^{-1}(z)$ for the source redshift
bin $l$ is approximately given by
\begin{equation}
\Sigma_{{\rm crit},l}^{-1}(z)\approx \frac{4\pi G}{c^2} D(z)\frac{D(z,\bar{z}_l)}{D(\bar{z}_l)},
\end{equation}
with $\bar{z}_l=(z_{l,{\rm min}}+z_{l,{\rm  max}})/2$. Given multiple
source and lens redshift bins, Equation~(\ref{eq:del2kap}) reduces to
a system of linear equations, which can be inverted easily to obtain
${\boldsymbol \delta}$ from lensing observables. In practice, however,
three-dimensional mass reconstruction is very noisy even with the high
source galaxy density of the HSC survey, and therefore an additional
regularization is essential. Here we adopt the Wiener filtering which
efficiently reduces the noise in the Fourier domain \citep{simon09}. We
assume that the noise is dominated by the shot noise. Then the noise
power between the $l$-th and $m$-th source redshift bins is given by
\begin{equation}
N_{lm}=\delta_{lm}\frac{\sigma_e^2}{\bar{n}_l},
\end{equation}
where
$\delta_{lm}$ is the Kronecker delta,
$\sigma_e$ is the root-mean-square of the source galaxy
ellipticity, and $\bar{n}_l$ is the mean number density of source
galaxies in the $l$-th bin, both of which are directly estimated from
the observation. On the other hand, the signal power in the
$k$-th and $n$-th lens redshift bins is given by 
\begin{equation}
S_{kn}=\delta_{kn}C_\ell(z_k),
\end{equation}
\begin{equation}
  C_\ell(z_k)=\frac{1}{(\Delta \chi_k)^2}\int_{z_{k,{\rm min}}}^{z_{k,{\rm max}}}
  dz\frac{P^{\rm m}(k=\ell/\chi)}{H(z)\chi^2},
\end{equation}
where
$\Delta \chi_k \approx \Delta z_k/H(\bar{z}_k)$,
$\Delta z_k =z_{l,{\rm max}}-z_{l,{\rm min}}$, and $P^{\rm m}(k)$ is
the matter power spectrum, which is computed using the halofit model
\citep{smith03,takahashi12}. The use of this signal power corresponds
to the transverse Wiener filter in \citet{simon09}. Given the expected
signal and noise powers, the three-dimensional mass reconstruction with
Wiener filtering from the observed (pixelized) shear maps in different
source redshift bins, ${\boldsymbol \gamma}$, is expressed as
\begin{equation}
{\boldsymbol \delta}({\boldsymbol \ell})=\tilde{W}(\ell)D^*({\boldsymbol \ell})
\left[\alpha \mathbf{S}^{-1}+\mathbf{Q}^{\rm T}\mathbf{N}^{-1}\mathbf{Q}\right]^{-1}
\mathbf{Q}^{\rm T}\mathbf{N}^{-1}{\boldsymbol \gamma}({\boldsymbol \ell}),
\label{eq:3dreconst}
\end{equation}
where $D({\boldsymbol \ell})={\boldsymbol \ell}^2/\ell^2$ \citep{kaiser93},
and $\tilde{W}(\ell)$ is the Fourier transform of the Gaussian
smoothing kernel (equation~\ref{eq:Gaussian}).

The parameter $\alpha$ in equation~(\ref{eq:3dreconst}) is an
important parameter which tunes the strength of the Wiener
filtering. The larger value of $\alpha$ leads to better
signal-to-noise ratios, although it also induces a bias in the
redshift of the reconstructed matter structure \citep{simon09}.
We try several different values of $\alpha$, and based
  on the trial result,
in this paper we adopt $\alpha=0.03$, which appears to represent a good
compromise between the signal-to-noise ratio and small bias in the
redshift. 

We need a large smoothing to reduce the shot noise in the
three-dimensional mass reconstruction. We adopt the pixel size of
$1'$, and the smoothing size of $\theta_{\rm s}=20'$ throughout this
section. We consider the source redshift range of $0.1<z<2.9$ with the
bin size of $\Delta z=0.1$, and the lens redshift range of
$0.05<z<1.05$ with the bin size of $\Delta z=0.1$.

\begin{figure*}
 \begin{center}
  \includegraphics[width=8.3cm]{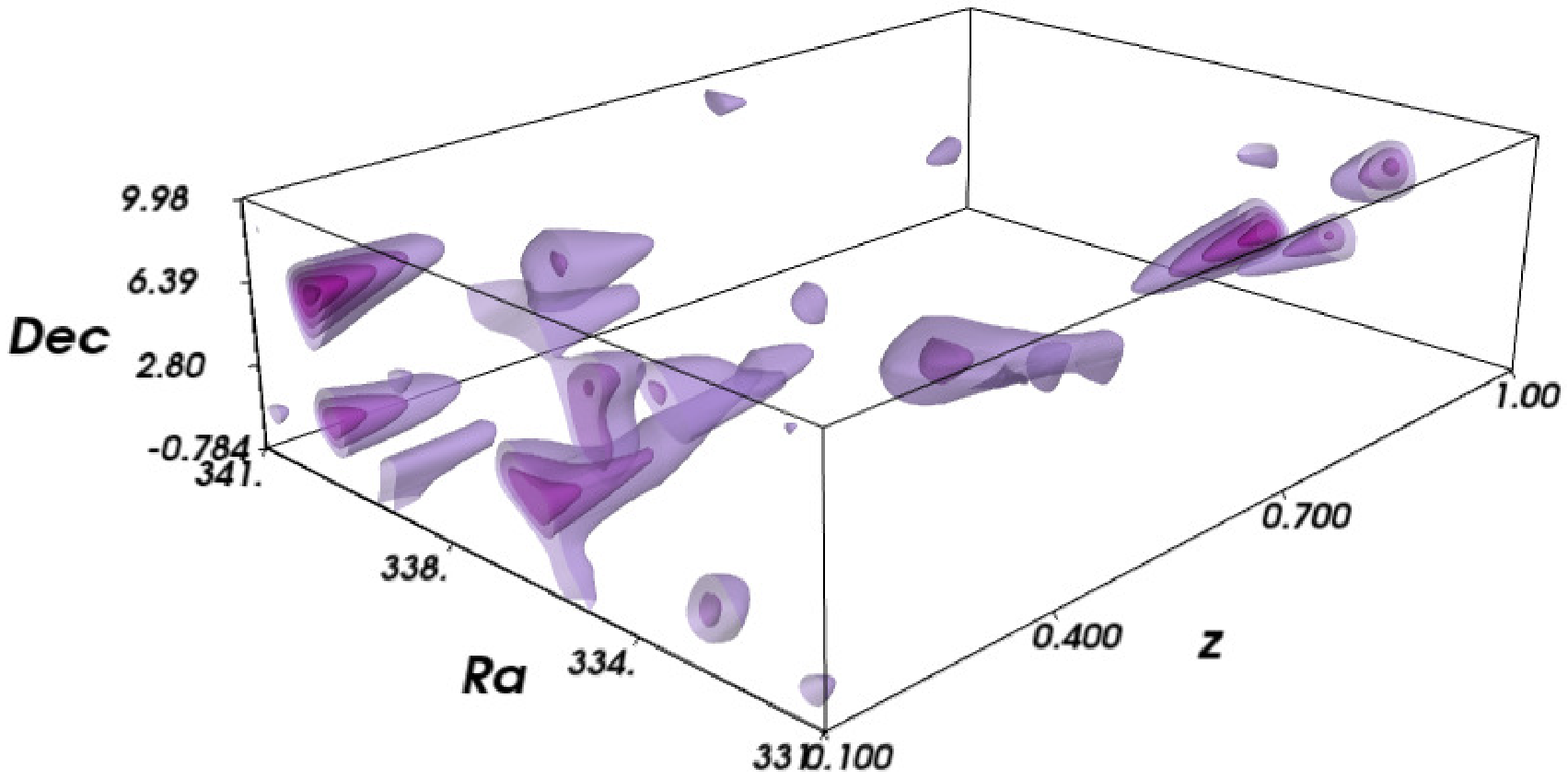} 
  \includegraphics[width=8.3cm]{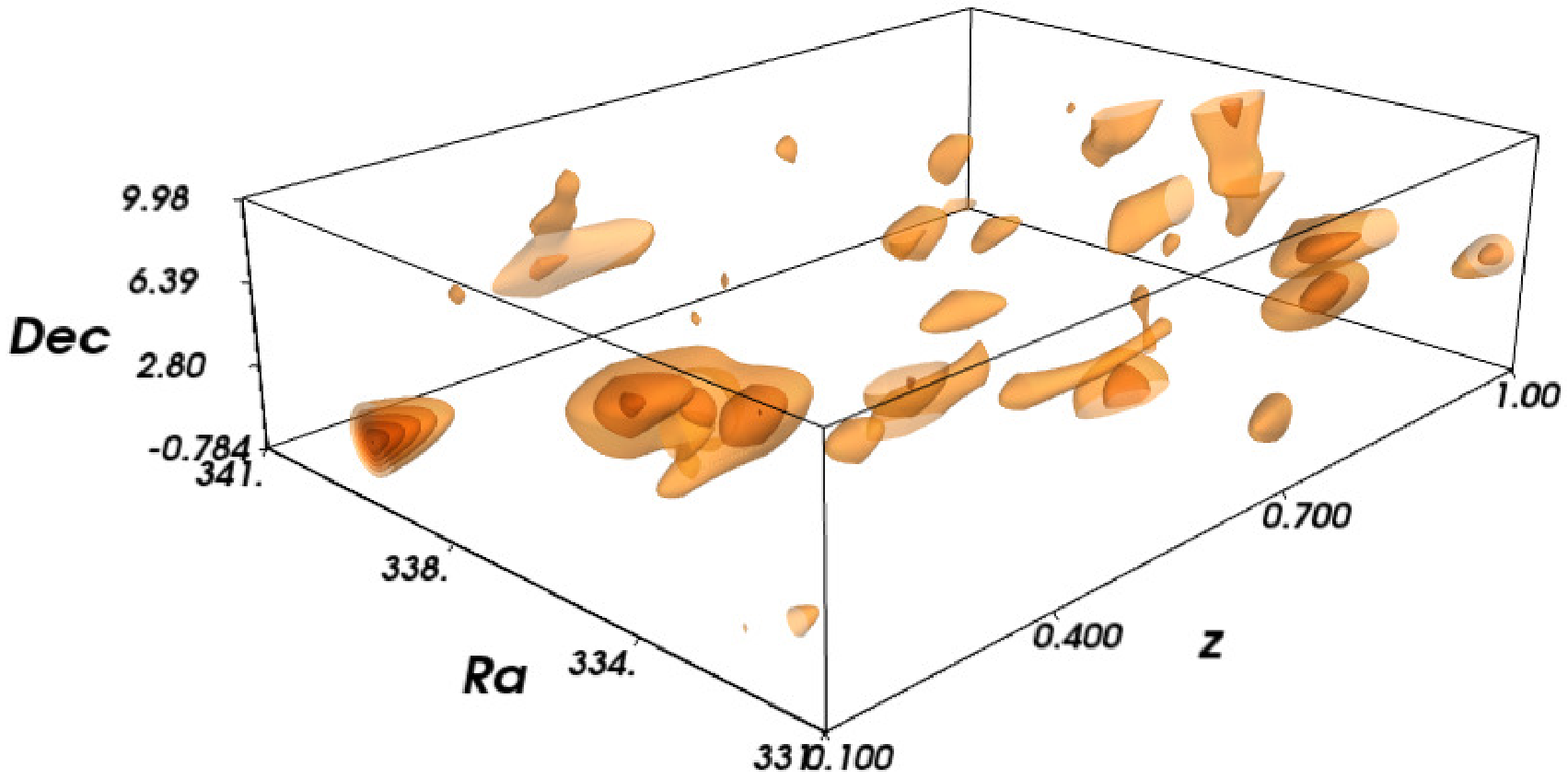} 
 \end{center}
\caption{Three-dimensional mass map from the VVDS region ({\it
    Left}). We also show the corresponding three-dimensional galaxy
  mass map from the photometric LRG sample ({\it Right}). The contours
  are drawn from 2$\sigma$ to 6$\sigma$
  with the 1$\sigma$ interval,
  where $\sigma$ is the rms map value.}
\label{fig:map3d} 
\end{figure*}

We show an example of reconstructed three-dimensional mass maps in
Figure~\ref{fig:map3d}. 

\subsection{Three-dimensional galaxy mass maps}\label{sec:galmap3d}

Three-dimensional galaxy mass maps are also constructed from the LRG
sample presented in Section~\ref{sec:galcat}. The stellar mass density
of each pixel with the side length of $1'$ and $\Delta z=0.1$ is
simply computed as $\rho_k=\sum_k M_{*,k}/V$, where $V$ is the volume of the
pixel. We then apply the same Gaussian smoothing kernel as used in the
three-dimensional mass reconstruction in Section~\ref{sec:wlmap3d}.
For each redshift slice, we again subtract the mean value.

An example of three-dimensional galaxy mass maps is also shown in
Figure~\ref{fig:map3d}.  

\subsection{Cross-correlation results}

Following Section~\ref{sec:cc2d}, we quantify the correlation between
the three-dimensional mass map (Section~\ref{sec:wlmap3d}) and the
three-dimensional galaxy mass map (Section~\ref{sec:galmap3d}) using
the Pearson correlation coefficient defined in equation~(\ref{eq:pearson}).
We cross-correlate mass maps in the same or different redshift bins.
In order to increase the signal-to-noise ratio further, we combine two
redshift bins to have 5 redshift slices at $0.05<z<1.05$ with the width
$\Delta z=0.2$. Thus, for each $E$- or $B$-mode mass map, we compute
$5\times 5=25$ correlation coefficients to check whether we successfully
recover the three-dimensional matter structure with weak lensing.
In the same manner as in the two-dimensional mass maps, we combine
results for all the 6 HSC S16A patches.

\begin{figure}
 \begin{center}
  \includegraphics[width=8cm]{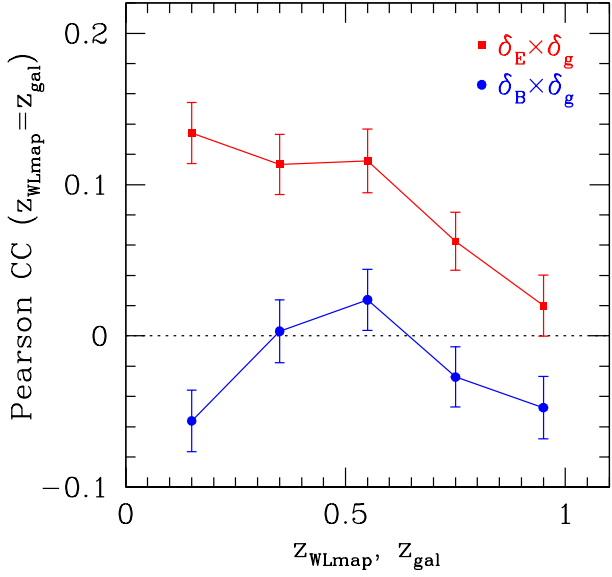} 
 \end{center}
\caption{Pearson correlation coefficients (equation~\ref{eq:pearson})
  between the three-dimensional mass maps from weak lensing and
  three-dimensional galaxy mass maps from LRGs. Here we show the diagonal
  correlation coefficients (i.e., same redshift bins for mass
  maps and galaxy mass maps) as a function of redshift. Both $E$-mode
  ({\it filled squares}) and $B$-mode ({\it filled circles}) mass map
  results are shown. Errors are estimated from 50 mock samples
  of the weak lensing shear catalog. }
\label{fig:plot_cov} 
\end{figure}

\begin{figure}
 \begin{center}
  \includegraphics[width=8cm]{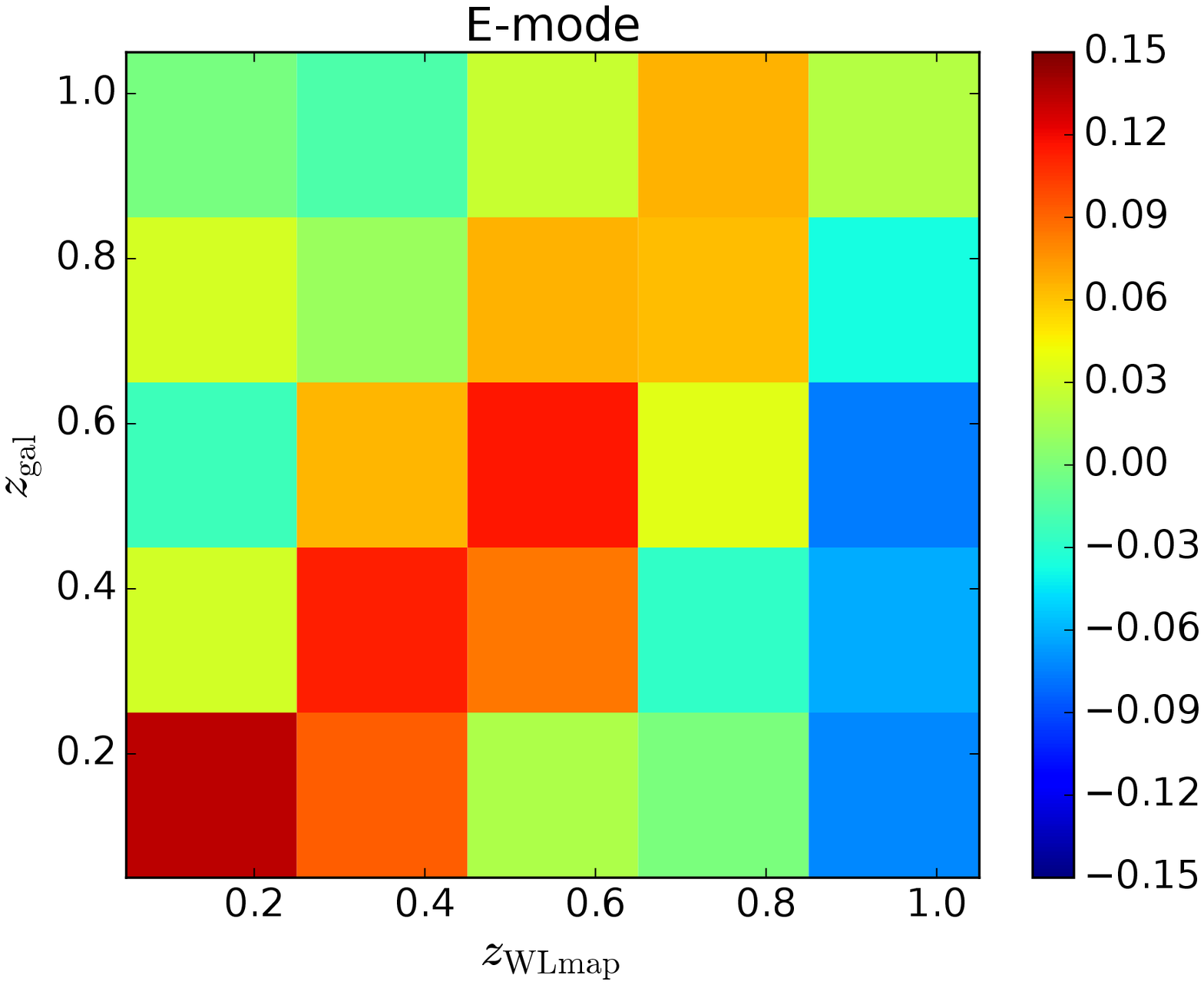} 
  \includegraphics[width=8cm]{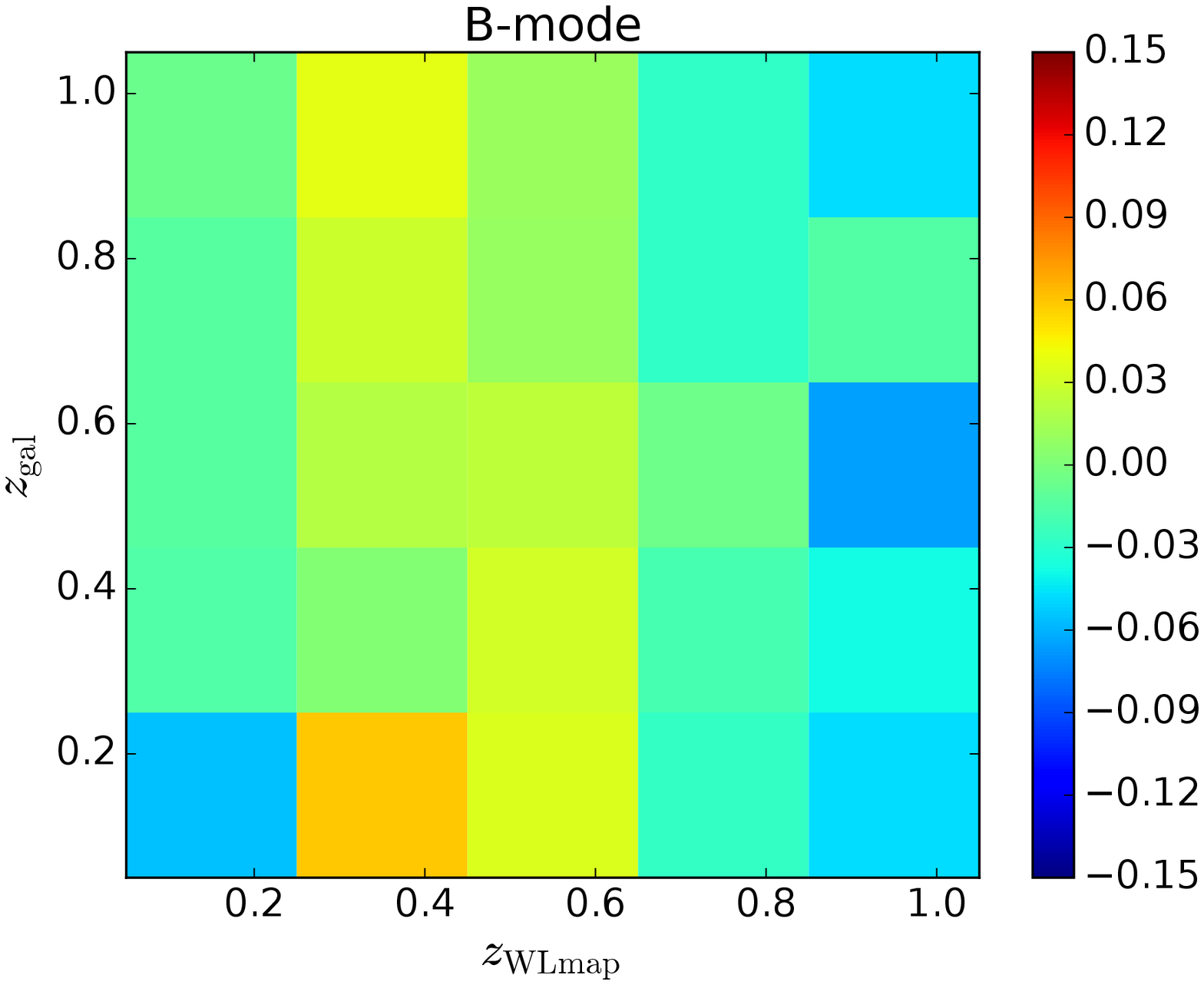} 
 \end{center}
\caption{Matrix of Pearson correlation coefficients for the same and
  different redshift bins between three-dimensional $E$-mode ({\it
    upper}) and $B$-mode ({\it lower}) mass maps and three-dimensional
  galaxy mass maps. The typical statistical error on the correlation
  coefficient is $\sim 0.02$.} 
\label{fig:plot_covmat} 
\end{figure}

Figure~\ref{fig:plot_cov} shows the diagonal part of the correlation
coefficients. We find that the cross-correlations are significantly
detected particularly at low redshifts, $z\lesssim 0.6$, which
indicates the successful three-dimensional mass reconstruction. 
The correlation coefficients are not very large due to
the large effect of the shot noise in three-dimensional mass
reconstruction. 
Although the $E$-mode correlation coefficients decrease at higher
redshifts, we find that this is partly due to the redshift bias in
reconstructed mass maps. This is obvious from
Figure~\ref{fig:plot_covmat}, which shows the full correlation matrix
of the three-dimensional mass maps and three-dimensional galaxy mass
maps. The Figure indicates that the three-dimensional mass
reconstruction is indeed successful, but the redshift of the
reconstructed mass distribution is biased low particularly at high
redshift. As discussed in Section~\ref{sec:wlmap3d}, this is in fact
expected in our mass reconstruction method because the Wiener
filtering method used here is a biased estimator. The redshift bias of
$\Delta z\sim 0.2$ at $z\sim 0.9$ for our choice of the parameter
$\alpha=0.03$ is in good agreement with the expected bias estimated by
\citet{simon09}.
Since the bias is understood fairly well, in principle
  we can construct the galaxy mass map with the expected redshift bias
  of the Wiener filtering method to make a fairer comparison, which we
  leave for future work.
Thus we conclude that we successfully reconstructed
three-dimensional mass map out to high redshift, which is made
possible thanks to the high number density of the weak lensing shape
catalog in the HSC survey. 

\section{Summary}\label{sec:summary}

We have presented weak lensing mass maps from the HSC S16A dataset
covering 167~deg$^2$. We have cross-correlated projected
two-dimensional mass maps with two-dimensional galaxy mass maps
constructed from stellar masses of photometric LRGs that are also
selected from the HSC data. We have found that the $E$-mode mass maps
correlate with the galaxy mass maps significantly, even with relatively
small smoothing sizes of $\theta_s=2'$. More specifically, the
cross-correlation coefficients are $\rho=0.54\pm0.03$ for
$\theta_s=8'$ and $\rho=0.34\pm 0.01$ for $\theta_s=2'$. This finding
confirms the validity of our weak lensing measurements and weak
lensing mass reconstructions. We have also checked for potential
systematic effects in mass maps by cross-correlating the weak lensing
mass maps with maps of various parameters that can be a source of
systematic effects in mass maps, and found that the cross-correlations
are sufficiently small. Finally, we reconstructed three-dimensional
mass maps from weak lensing using photometric redshift measurements of
individual source galaxies. We have found that the three-dimensional
mass map correlates reasonably well with three-dimensional galaxy mass
map, which indicates that our three-dimensional weak lensing mass
reconstruction is successful.

Our work demonstrates the power of the HSC survey for weak lensing
studies. This is mainly due to the high number density of source
galaxies of $\bar{n}\sim 25$~arcmin$^{-2}$ for weak lensing analysis. In
particular, previous three-dimensional weak lensing mass reconstructions
have been limited to relatively small areas \citep[e.g.,][]{massey07},
and this work successfully applied the technique to much wider area to
obtain wide-field three-dimensional mass maps. Given the validation of
mass maps presented in this paper, we plan to use HSC weak lensing
mass maps to study the large-scale structure of dark matter and
baryons, including the construction of a mass-selected cluster sample
\citep{miyazaki17b} and the correlation of dark matter and hot gas
from the cross-correlation of weak lensing mass maps and
Sunyaev-Zel'dovich maps \citep{osato17}.

\begin{ack}
We thank T. Hamana and K. Osato for useful discussions, and the
anonymous referee for useful comments.  
This work was supported in part by World Premier International Research Center Initiative (WPI Initiative), MEXT, Japan, and JSPS KAKENHI Grant Number 26800093, 15H05887, 15H05892, and 15H05893.
MO acknowledges financial support from JST CREST Grant Number JPMJCR1414.
RM is supported by the US Department of Energy Early Career Award Program.
HM is supported by the Jet Propulsion Laboratory, California Institute of Technology, under a contract with the National Aeronautics and Space Administration.

The Hyper Suprime-Cam (HSC) collaboration includes the astronomical communities of Japan and Taiwan, and Princeton University.  The HSC instrumentation and software were developed by the National Astronomical Observatory of Japan (NAOJ), the Kavli Institute for the Physics and Mathematics of the Universe (Kavli IPMU), the University of Tokyo, the High Energy Accelerator Research Organization (KEK), the Academia Sinica Institute for Astronomy and Astrophysics in Taiwan (ASIAA), and Princeton University.  Funding was contributed by the FIRST program from Japanese Cabinet Office, the Ministry of Education, Culture, Sports, Science and Technology (MEXT), the Japan Society for the Promotion of Science (JSPS),  Japan Science and Technology Agency  (JST),  the Toray Science  Foundation, NAOJ, Kavli IPMU, KEK, ASIAA,  and Princeton University.

The Pan-STARRS1 Surveys (PS1) have been made possible through contributions of the Institute for Astronomy, the University of Hawaii, the Pan-STARRS Project Office, the Max-Planck Society and its participating institutes, the Max Planck Institute for Astronomy, Heidelberg and the Max Planck Institute for Extraterrestrial Physics, Garching, The Johns Hopkins University, Durham University, the University of Edinburgh, Queen's University Belfast, the Harvard-Smithsonian Center for Astrophysics, the Las Cumbres Observatory Global Telescope Network Incorporated, the National Central University of Taiwan, the Space Telescope Science Institute, the National Aeronautics and Space Administration under Grant No. NNX08AR22G issued through the Planetary Science Division of the NASA Science Mission Directorate, the National Science Foundation under Grant No. AST-1238877, the University of Maryland, and Eotvos Lorand University (ELTE).
 
This paper makes use of software developed for the Large Synoptic Survey Telescope. We thank the LSST Project for making their code available as free software at http://dm.lsst.org.

Based in part on data collected at the Subaru Telescope and retrieved from the HSC data archive system, which is operated by the Subaru Telescope and Astronomy Data Center at National Astronomical Observatory of Japan.
\end{ack}

\section*{Appendix 1. Mock shear catalogs}
We construct mock shear catalogs which take full account of the survey
geometry, the spatial inhomogeneity, and the redshift distribution of
galaxies. We do so by adopting a real shear catalog from the
observations and replacing the ellipticities of individual galaxies with
mock ellipticity values that include the cosmic shear from
ray-tracing simulations. 

First we review the relation between the ellipticity and the shear in
our shear catalog. In this paper we adopt the re-Gaussianization
method \citep{hirata03}, which uses the second moments of the surface 
brightness distribution of the source, 
$Q_{ij}\propto \int d\vec{\theta} I(\vec{\theta})\theta_i\theta_j$,
where $I(\vec{\theta})$ is the surface brightness distribution and the
coordinate origin is set to the center of the source, to define the
ellipticity of each galaxy. Specifically, a complex ellipticity is
defined as $\epsilon=(Q_{11}-Q_{22}+2iQ_{12})/(Q_{11}+Q_{22})$.
In practice a weight function is included in the measurement of the
second moment, which is ignored here just for simplicity.
The intrinsic ellipticity $\epsilon^{\rm int}$ and the observed
ellipticity with a weak lensing effect $\epsilon^{\rm lens}$ are
related as \citep[e.g.,][]{seitz95} 
\begin{equation}
\epsilon^{\rm lens}=\frac{\epsilon^{\rm int}+2g+g^2\epsilon^{\rm
    int,*}}{1+|g|^2+2{\rm Re}[g\epsilon^{\rm int, *}]},
\label{eq:lensellip}
\end{equation}
where $g=\gamma/(1-\kappa)$ is the so-called reduced shear.

In order to construct a mock shear catalog, we adopt the real shear
catalog from the HSC observation. In the mock catalog, the coordinates
of all the galaxies in the shear catalogs are kept unchanged, but we
simply replace the observed ellipticities of the individual galaxies,
$\epsilon^{\rm obs}$, with simulated values. To derive simulated
ellipticity values, first we randomly rotate each galaxy,
$\epsilon^{\rm ran}=e^{i\phi}\epsilon^{\rm obs}$, where $\phi$ is a
random number between $0$ and $2\pi$. We need to distinguish the
intrinsic ellipticity from the measurement error because they have
different impacts on the shear responsivity. For each galaxy, the
shear catalog has an estimate of the intrinsic rms ellipticity 
$\sigma_{\rm int}$ (parameter {\tt
  ishape\_hsm\_regauss\_derived\_rms\_e}) and the measurement
error $\sigma_{\rm sta}$ (parameter {\tt ishape\_hsm\_regauss\_derived\_sigma\_e}).
For each galaxy, we derive a randomized intrinsic ellipticity as
\begin{equation}
\epsilon^{\rm int}=f\epsilon^{\rm ran},
\end{equation}
\begin{equation}
f=\frac{\sigma_{\rm int}}{\sqrt{\sigma_{\rm int}^2+\sigma_{\rm sta}^2}}.
\end{equation}
We then add weak lensing effects via equation~(\ref{eq:lensellip}) to
convert $\epsilon^{\rm int}$ to lensed galaxy ellipticity
$\epsilon^{\rm lens}$. For weak lensing, we take all-sky weak lensing
maps presented in \citet{takahashi17}. The cosmological model
is from the best-fit result of the Wilkinson Microwave Anisotropy Probe
nine year data \citep{hinshaw13} with $\Omega_M=0.279$,
$\Omega_b=0.046$, $\Omega_\Lambda=0.721$, $h=0.7$, $n_s=0.97$, and
$\sigma_8=0.82$. \citet{takahashi17} created all-sky weak
lensing maps at 38 source redshift slices from $z=0.05$ to $5.3$,
which are stored in a HEALPix format \citep{gorski05}. Although there
are realizations with different angular resolutions, we use a low
resolution version with NSIDE equal to 4096, which roughly corresponds
to a pixel size of $\sim 1$~arcmin. For each galaxy, we randomly
assign its redshift following the photometric redshift PDF of that
galaxy \citep[see][]{tanaka17}, and obtain values of convergence
$\kappa$ and complex shear $\gamma$ from the all-sky weak lensing maps
at two adjacent redshift slices and linearly interpolating map
values. The value of $\gamma$ is rescaled by a factor of $(1+m)$ for
each galaxy in order to account for the multiplicative bias 
\citep[see][]{mandelbaum17}. After adding weak lensing effects from  
the all-sky ray-tracing simulations, we add a random measurement
noise, $\epsilon^{\rm mock}=\epsilon^{\rm lens}+(N_1+iN_2)$, where
$N_i$ is a random value drawn from a normal distribution with a
standard deviation of $\sigma_{\rm sta}$. From this procedure we
create a list of mock ellipticities $\epsilon^{\rm mock}$ for the weak
lensing shear catalog, which properly include the effect of the cosmic
shear. When generating different realizations, we randomly rotate the
all-sky weak lensing map before assigning weak lensing effects to
randomized galaxies so that we take the different realization of the
cosmic shear from the different patch of the all-sky weak lensing map.


\end{document}